\pdfoutput=1
    \documentclass[sigconf, authorversion=true]{acmart}
 \usepackage{multirow,acmart-taps}
 \usepackage{bigdelim} %% Might make problems with ACM
\graphicspath{{figures/}}
 \usepackage{subcaption}
 \usepackage{tikz,pgf} %% Might make problems with ACM
 \usepackage{pgfplots} %% Might make problems with ACM
\pgfplotsset{compat=1.17}
\usetikzlibrary{arrows, automata, calc, positioning,shapes}
\usetikzlibrary{decorations.pathreplacing, calligraphy}
 \usetikzlibrary{patterns}
 \usetikzlibrary{backgrounds}
 \pgfarrowsdeclarecombine*{spaced stealth'}{spaced stealth'}{stealth'}{stealth'}{space}{space}
 \tikzset{>=spaced stealth'}
 \tikzset{footnotesize/.style = {font=\footnotesize}}
 \usepgfplotslibrary{external}
\definecolor{dkgreen}{rgb}{0,0.6,0}
\definecolor{gray}{rgb}{0.5,0.5,0.5}
\definecolor{mauve}{rgb}{0.58,0,0.82}
\definecolor{linkcolor}{rgb}{0.65,0,0}
\definecolor{citecolor}{rgb}{0,0.4,0}
\definecolor{urlcolor}{rgb}{0,0,0.65}
\definecolor{nicegreenfill}{RGB}{213,232,212}
\definecolor{nicegreenborder}{RGB}{151,192,128}
\definecolor{niceorangefill}{RGB}{255,230,204}
\definecolor{niceorangeborder}{RGB}{225,178,61}
\definecolor{niceyellowfill}{RGB}{255,242,204}
\definecolor{niceyellowborder}{RGB}{222,195,114}
\definecolor{niceredfill}{RGB}{171,70,70}
\definecolor{niceredborder}{RGB}{150,38,38}
\definecolor{nicebluefill}{RGB}{218,232,252}
\definecolor{niceblueborder}{RGB}{133,161,203}
\definecolor{plotblue}{rgb}{0.00000,0.44700,0.74100}%
\definecolor{plotorange}{rgb}{0.85000,0.32500,0.09800}%

\usepackage[vlined, linesnumbered]{algorithm2e}
\usepackage{listings}
\SetKwProg{Fn}{Function}{}{}
\SetKwInOut{Input}{input}
\SetKwInOut{Output}{output}
\SetKwFunction{evicts}{evicts}
\SetKwFunction{constructEvset}{constructEvset}
\SetKwFunction{evsetFinding}{evsetFinding}
 \usepackage{units}
    \newcommand{\revision}[1]{{#1}}
\setcopyright{acmcopyright}
\copyrightyear{2023}
\acmYear{2023}
\setcopyright{acmlicensed}\acmConference[ASIA CCS '23]{ACM ASIA Conference on Computer and Communications Security}{July 10--14, 2023}{Melbourne, VIC, Australia}
\acmBooktitle{ACM ASIA Conference on Computer and Communications Security (ASIA CCS '23), July 10--14, 2023, Melbourne, VIC, Australia}
\acmPrice{15.00}
\acmDOI{10.1145/3579856.3582838}
\acmISBN{979-8-4007-0098-9/23/07}

\begin{document}

%%%%%%%%%%%%%%%%%%%%%%%%%%%%%%%%%%%%%%%%%%%%%%%%%%%%%%%%
%% Define title and authors
%%%%%%%%%%%%%%%%%%%%%%%%%%%%%%%%%%%%%%%%%%%%%%%%%%%%%%%%

%%
%% The "title" command has an optional parameter,
%% allowing the author to define a "short title" to be used in page headers.
\title[IOTLB-SC]{IOTLB-SC: An Accelerator-Independent Leakage Source in Modern Cloud Systems}

%%
%% The "author" command and its associated commands are used to define
%% the authors and their affiliations.
%\ifdefined\PUBLICATION
    \author{Thore Tiemann}
    \email{t.tiemann@uni-luebeck.de}
    \orcid{0000-0001-9018-4226}
    \affiliation{%
        \institution{University of Lübeck}
        \streetaddress{Ratzeburger Allee 160}
        \city{Lübeck}
        \state{SH}
        \country{Germany}
        \postcode{23562}
    }
    \author{Zane Weissman}
    \email{zweissman@wpi.edu}
    \orcid{0000-0002-8594-5390}
    \affiliation{%
        \institution{Worcester Polytechnic Institute}
        \streetaddress{100 Institute Road}
        \city{Worcester}
        \state{MA}
        \country{USA}
        \postcode{01609-2280}
    }
    \author{Thomas Eisenbarth}
    \email{thomas.eisenbarth@uni-luebeck.de}
    \orcid{0000-0003-1116-6973}
    \affiliation{%
        \institution{University of Lübeck}
        \streetaddress{Ratzeburger Allee 160}
        \city{Lübeck}
        \state{SH}
        \country{Germany}
        \postcode{23562}
    }
    \author{Berk Sunar}
    \email{sunar@wpi.edu}
    \orcid{0000-0001-5404-5368}
    \affiliation{%
        \institution{Worcester Polytechnic Institute}
        \streetaddress{100 Institute Road}
        \city{Worcester}
        \state{MA}
        \country{USA}
        \postcode{01609-2280}
    }
    %%
    %% By default, the full list of authors will be used in the page
    %% headers. Often, this list is too long, and will overlap
    %% other information printed in the page headers. This command allows
    %% the author to define a more concise list
    %% of authors' names for this purpose.
    \renewcommand{\shortauthors}{Tiemann et al.}
%\fi

\begin{abstract}
%% Abstract:
Hardware peripherals such as GPUs and FPGAs are commonly available in server-grade computing to accelerate specific compute tasks, from database queries to machine learning. CSPs have integrated these accelerators into their infrastructure and let tenants combine and configure these components flexibly, based on their needs.
Securing I/O interfaces is critical to ensure proper isolation between tenants in these highly complex, heterogeneous, yet shared server systems, especially in the cloud, where some peripherals may be under control of a malicious tenant.

In this work, we investigate the interfaces that connect peripheral hardware components to each other and the rest of the system. We show that the I/O memory management units (IOMMUs) --- intended to ensure proper isolation of peripherals --- are the source of a new attack surface: the I/O translation look-aside buffer (IOTLB). We show that by using an FPGA accelerator card one can gain precise information over IOTLB activity. That information can be used for covert communication between peripherals without bothering CPU or to directly extract leakage from neighboring accelerated compute jobs such as GPU-accelerated databases. 
We present the first qualitative and quantitative analysis of this newly uncovered attack surface \revision{before fine-grained channels become widely viable with the introduction of CXL and PCIe 5.0}. In addition, we propose possible countermeasures that software developers, hardware designers, and system administrators can use to suppress the observed side-channel leakages and analyze their implicit costs. 
\end{abstract}

%%
%% The code below is generated by the tool at http://dl.acm.org/ccs.cfm.
%% Please copy and paste the code instead of the example below.
%%


%%
%% Keywords. The author(s) should pick words that accurately describe
%% the work being presented. Separate the keywords with commas.

%%
%% This command processes the author and affiliation and title
%% information and builds the first part of the formatted document.
\settopmatter{printfolios=true}
\maketitle

%% a whole new intro by zane 1/31:
\section{Introduction}

% Story
%  Modern Cloud: Mix of Peripherals and CPU VMs: Peripherals used by CSP and by Tenants
%  IOMMU is essential boundary for memory protection
% In the past: focus on caches, as the only non-stoppable side channel in cloud environments %Bulk, hello from the other side, what else?
%  Cache and other microarchitectural timing attacks have long been known to be a problem
%  With rental of FPGA, unfettered control over interaction with environment, including IOMMU
%  Show that there are inherent security issues due to IOMMU- 

Modern server-grade computing infrastructures are becoming more heterogeneous: computational needs are spread over fast and flexible CPUs as well as powerful peripherals such as smart storage, GPUs, smart NICs and FPGAs. Major cloud service providers (CSPs) have started to shift tasks such as networking, memory management and VM management into more specialized hardware peripherals~\cite{alibaba2020introducing,amazon2022nitro,firestone2018azure}, freeing up precious CPU time that is rented to more tenants who share the same hardware. 
These multi-tenant, peripheral-heavy cloud systems rely on increasingly interlinked memory systems to provide high throughput for shared, scalable and parallelized cloud infrastructure.
Technologies like VT-d, DDIO, \revision{and CXL} allow peripherals to not only directly read and write to the memory of a virtual machine, but to also use a CPU's shared cache to speed up repeated reads and writes.

On a logic layer, input-output memory management units (IOMMUs) enforce memory isolation between these peripherals and guest VMs running on CPUs, making IOMMUs a key component for ensuring security of the cloud infrastructure~\cite{amd2021iommu,intel2019vtd,markettos2019thunderclap}.
The IOMMU ensures that accesses to virtual memory spaces are isolated and appropriately virtualized: e.g., devices may handle only I/O-specific virtual addresses and not the CPU-side virtual addresses or the underlying system's physical addresses; in addition, devices may only access memory with the appropriate permissions set.

However, when many tenants share the same hardware, side effects in these complex shared memory systems weaken the security promises of virtualization that make highly scalable multi-tenant cloud computing possible.
These side effects of shared hardware are exploited by microarchitectural attacks, most prominently cache attacks. 
Cache attacks exploit the measurable difference in access times to the many tiers of modern caches to overcome the sophisticated memory isolation mechanisms that protect tenants' data and computation from each other.
Besides cache attacks, which have been successfully applied in commercial cloud settings~\cite{inci2016cache,maurice2017hello}, microarchitectural attacks like Meltdown~\cite{lipp2018meltdown} and related MDS attacks~\cite{vanschaik2019ridl,Schwarz2019ZombieLoad,borello2022aepic} as well as Rowhammer attacks pose a real threat in shared cloud environments. 
One malicious tenant may, after successful co-location~\cite{ristenpart2009hey,zhang2014cross}, use these microarchitectural side effects to glean sensitive information from co-located VMs.

While Meltdown and other MDS-style attacks have mostly been patched with microcode updates~\cite{mitre2018msbds,mitre2018mlpds}, they often also require CSPs to disable simultaneous multi-threading between separate security domains for full protection.
Rowhammer attacks are significantly mitigated by usage of ECC memory and newer DDR4 and DDR5 architectures, even though vulnerability of both older ECC memory modules and newer ECC-free DDR4 memory to Rowhammer has been practically verified~\cite{cojocar2019exploiting,pessl2016drama,frigo2020trrespass}.

Cache attacks, however, are much more difficult to prevent, as the contention and timing differences that enable attacks such as Prime+Probe are inherent to modern cache architecture~\cite{osvik2006cache}. 
Many system-level solutions like cache partitioning have been proposed~\cite{ye2014coloris,sanchez2011vantage}, but have not been widely implemented in hardware and are costly in performance to implement in firmware or hypervisors.
The most common way to prevent these attacks is through constant-time implementation of security-critical code; that is, rather than removing the leakage channel inherent to the cloud architecture, software developers must make sure their code does not leak sensitive data on timing-based channels~\cite{almeida2016verifying,barthe2014system,blazy2017verifying,wichelmann2018microwalk}. 

While most research in microarchitectural attacks has focused on attacks from core to core on CPUs, caches are no longer only accessible by CPUs.
Intel's DDIO technology, present on all recent Intel server architectures, allows high speed peripherals to directly access a CPU's shared cache without interrupting CPU execution~\cite{intel2012ddio}.
Cloud users may rent peripherals such as purpose-specific GPU or FPGA cloud instances for higher performance in particular workloads.
In such heterogeneous compute environments, security is even more challenging, as tenants are no longer confined to virtual machines (VMs) on the CPU, but may additionally have control over peripherals. 
With CSPs renting instances that grant tenants full access to FPGAs designed specifically for heterogeneous computation~\cite{amazon2019ec2f1,alibaba2019fpgacloud,xilinx2017aws}, it becomes trivial for attackers to gain sufficient control over peripherals in the cloud that are more than capable of exploiting microarchitectural vulnerabilities.

First works have used peripherals like network cards~\cite{kurth2020netcat} and FPGAs~\cite{weissman2020jackhammer,purnal2022double} to target CPU caches as a powerful shared resource that is accessible by VMs and peripherals alike. 
These works indicate that not only are cache attacks mounted from peripherals possible; they can leak information about the private operations of both CPUs \emph{and} peripherals. Furthermore, classical cache attack methods can become more powerful when the attacker controls peripherals in addition to a VM on the same machine.
As of now, several other components that are also shared by peripherals such as the IOMMU, which is the main line of defense against compromised peripherals, remain unstudied and may open up new attack surfaces.

\paragraph{Our Contribution}
This work exposes a vulnerability in an overlooked attack surface present in multi-tenant, peripheral-heavy cloud systems: the microarchitecture of the I/O Memory Management Unit (IOMMU).
Knowing that the IOMMUs in modern CPUs have translation look-aside buffers (IOTLBs) to speed up repeated translations~\cite{amd2021iommu,intel2019vtd,neugebauer2018pcieperformance}, we present a hardware design for an FPGA acceleration card that uses memory access timing to reliably identify whether or not a translation is present in an IOTLB.
With that design, we propose and evaluate an algorithm for IOTLB eviction set finding. With those eviction sets, we demonstrate the first two IOTLB-based covert channels.
We use the FPGA to collect side-channel IOTLB traces from two other peripheral devices and analyze the viability and threat models of a full side-channel attack.

We show that the IOTLB is the source of side-channel vulnerability that CSPs are currently not aware of and thus do not protect against. We show that the IOTLB is an excellent source for constructing covert channels between co-located peripherals and can also be abused to extract information from neighboring peripherals such as GPU-accelerated databases. 
We provide comprehensive threat analysis of this vulnerability\revision{, in both the present and the near future,} and present viable defenses and countermeasures.
In summary, our main contributions are:
\begin{itemize}
  \item We demonstrate a \emph{previously ignored IOTLB timing side-channel} against PCIe peripherals \revision{\emph{before} technologies such as CXL and PCIe 5.0 gain widespread adoption, and fine-grained attacks become viable on a large installation base}.
  \item We \revision{develop} a new algorithm that finds eviction sets without any prior assumptions of \revision{organization and demonstrate its advantages in finding IOTLB eviction sets over a similar eviction set finding algorithm.}
  \item We use a custom FPGA hardware function to exploit the IOTLB timing side-channel and study traces collected from an SQL database acceleration library for a GPU.
  \item We leak IOTLB timing side-channel traces from a GPU-accelerated SQL database library and analyze the vulnerability of the library to a practical attack.
  \item We demonstrate the \emph{first} two IOTLB covert channels, including a peripheral-to-peripheral channel with a generic application as the sender and our custom FPGA function as the receiver.
  \item We propose countermeasures for applications, cloud systems, and IOMMU implementations to counter the side-channel we identified.
\end{itemize}

%%%%%%%%%%%%%%%%%%%%%%%%%%%%%%%%
\section{Background}
\label{sec:background}
%%%%%%%%%%%%%%%%%%%%%%%%%%%%%%%%
When multiple hardware resources share data, it is often desirable to have direct memory access (DMA) from one resource to another.
However, simply allowing any peripheral to read or write a host CPU's memory would be disastrous for security, especially in virtualized environments with multiple users sharing the CPU.
AMD’s AMD-Vi and Intel’s VT-d features (present on both companies' performance desktop and server processors for the better part of a decade) allow for virtualized DMA with IOMMUs that dynamically map and translate virtual addresses used specifically by peripherals to access CPU memory. 
To speed up repeated access to the same memory location, IOMMUs often include translation look-aside buffers (TLBs, or IOTLBs when they are in IOMMUs) which cache recently translated I/O virtual addresses and their corresponding physical addresses to avoid the slow page-table walks otherwise required for translation.
Like CPU caches and TLBs, which perform a similar function for CPU memory accesses and address translations, IOTLBs introduce a timing-based side-channel vulnerability.

%%%%%%%%%%%%%%%%%%%%%%%%%%%%%%%%%%%%%%%%%%%%%%%%%%%%%%%
\subsection{Caches and TLBs}
\label{sec:caches-and-tlbs}
%%%%%%%%%%%%%%%%%%%%%%%%%%%%%%%%%%%%%%%%%%%%%%%%%%%%%%%
A cache stores data for faster access.
A translation look-aside buffer (TLB) is technically just another cache, though rather than caching the data or instructions stored at an address, it caches an address translation. However, throughout this paper we will refer to memory caches as simply ``caches''. 
Intel's documentation\cite{intel2016optimization} and several works reverse-engineering cache architectures~\cite{hund2013practical,irazoqui2015systematic,oren2015spy,liu2015last} and TLB architectures~\cite{gras2018translation,tatar2022tlbdr} reveal that TLBs on modern Intel CPUs are organized very similarly to modern CPU memory caches. Modern TLBs and caches are typically organized into \textbf{sets} and \textbf{ways}. The number of ways is the number of entries each set can contain. For TLBs, each virtual address is mapped to one set, but can occupy any way within that set. When a set is full, old entries may be evicted to make room for new ones. A set of addresses which reliably causes the eviction of all other entries in a set when accessed is called an \textbf{eviction set}. 
A minimal eviction set contains as many addresses as there are ways in the cache/TLB and therefore fills an entire cache set when accessed~\cite{vila2019theory}.

%%%%%%%%%%%%%%%%%%%%%%%%%%%%%%%%%%%%%%%%%
\subsection{Side-Channel Attacks}
\label{sec:sidechannels}
%%%%%%%%%%%%%%%%%%%%%%%%%%%%%%%%%%%%%%%%%
Timing side-channel attacks against the CPU's cache are widely studied and well understood: researchers have crafted several variants \cite{yarom2014flush,liu2015last,gruss2016flush,disselkoen2017prime,purnal2021primescope}, used them as part of more complicated microarchitectural attacks \cite{lipp2018meltdown,kocher2019spectre}, and built defenses against them \cite{gruss2017tsx,liu2016catalyst,ye2014coloris}.
There are many cache side-channel strategies that work in different memory-sharing scenarios and have quite varied temporal and address resolutions. These are two of the most common and useful attack techniques:

\emph{Flush+Reload} (F+R) \cite{yarom2014flush} requires shared memory between the attacker and the victim and has three steps:
\textbf{1)} The attacker flushes the cache line of interest.
\textbf{2)} She then waits for the victim to execute.
Later, \textbf{3)} she reloads the flushed line and measures the reload latency.
If the latency is low, the cache line was served from the cache hierarchy, so the cache line was accessed by the victim.

\emph{Prime+Probe} (P+P) does not require shared memory at the cost of a lower temporal resolution than F+R since the attacker checks the status of the cache by probing a whole cache set rather than flushing or reloading a single line.
P+P has three steps: 
\textbf{1)} The attacker primes the cache set under surveillance with dummy data by accessing a proper eviction set,
\textbf{2)} she waits for the victim to execute,
\textbf{3)} she accesses the eviction set again and measures the access latency (probing).
If the latency is above a certain threshold, some parts of the eviction set were evicted by the victim process, 
meaning that the victim accessed cache lines belonging to the cache set under surveillance~\cite{liu2015last}.

%%%%%%%%%%%%%%%%%%%%%%%%%%%%%%%%%%%%%%%%%%%%%%%%%%%%%%%%%
\subsection{Attacks on TLBs}
\label{sec:tlb-attacks}
%%%%%%%%%%%%%%%%%%%%%%%%%%%%%%%%%%%%%%%%%%%%%%%%%%%%%%%%%
In 1995, Silbert et al. remarked in a security analysis of Intel CPU architectures that "all 80x86 [now more commonly called x86] processors have a translation look-aside buffer (TLB) that [\ldots] has potential for use as a covert timing channel"~\cite{sibert1995intel}. In 2013, Hund et al.~\cite{hund2013practical} demonstrated that a TLB timing side-channel on then-modern Intel CPUs could reveal if a page was mapped by the operating system even if the user does not have permission to access the page directly. They demonstrated that this exploit could be used to identify the pages used by the kernel, even when the addresses of the pages were randomized (a common defense against side-channel attacks of many types). In 2017, Gras et al. crafted an attack that uses a cache side-channel to identify TLB evictions. This was a robust attack that can be mounted even from JavaScript to de-randomize kernel pages~\cite{gras2017aslr}. Gras et al.'s ``TLBleed''  in 2018~\cite{gras2018translation} showed that TLBs in modern Intel CPUs were vulnerable to timing side-channel attacks of the sort that are typically used on CPU memory caches, and can be used for similarly complex attacks: with the help of some machine learning, the TLB side-channels on Skylake, Broadwell, and Coffeelake CPUs can be used to recover a key from an Edward-curve cryptographic function.

%%%%%%%%%%%%%%%%%%%%%%
\subsection{PCIe}
\label{sec:pcie}
%%%%%%%%%%%%%%%%%%%%%%
Peripheral Component Interconnect Express (PCIe)~\cite{pcisig2006pcie} is the backbone of modern desktop and server systems.
While often referred to as a bus, PCIe uses a high-speed point-to-point topology with devices being connected to switches or directly to a root port via serial links.
The root complex connects the PCIe network to the CPU and the main memory.
On a PCIe network, all devices can send memory requests to each other and to the main memory.
An IOMMU can be used to virtualize addresses used by PCIe devices and to implement access restrictions.
If supported, each root port of a root complex may define access rules for inter-device communication and implement them in the PCIe switches.

Two recent works \cite{tan2021invisibleprobe,khaliq2021timing} describe covert- and side-channel attacks that rely on PCIe bus contention. A preliminary is that the two devices involved share the same PCIe switch. 
In contrast, our work assumes the two devices to share a PCIe root port. 
Our assumption is less restrictive as any two PCIe devices sharing a switch share a root port, but devices sharing a root port do not necessarily share a switch, as root ports can have many lanes to support multiple devices without sharing a physical bus \cite{pcisig2006pcie}.

\revision{
Currently, PCIe 3.0 is the prevailing PCIe specification for commodity hardware. 
After a short period of CPUs supporting PCIe 4.0, PCIe specification 5.0 is the upcoming standard for the next generations of server-grade CPUs. 
CPUs supporting PCIe 5.0 are scheduled for November 2022 and January 2023, respectively \cite{amd2022cxl,shilov2022cxl}. 
PCIe 5.0 doubles transfer rates compared to PCIe 4.0, making the interconnect compete with main memory speeds. 
As a result, PCIe 5.0 physical layer is also used by a new protocol named Compute eXpress Link (CXL) \cite{sharma2019cxl}. 
CXL supports three sub-protocols: \emph{CXL.io} is based on PCIe and enables CXL devices to share the PCIe infrastructure with PCIe devices unaware of CXL. 
With \emph{CXL.cache}, devices are enabled to cache data from main memory while maintaining coherency between the main memory, the CPU caches and the accelerator cached copy. 
\emph{CXL.mem} is used by a host CPU to access CXL device memory and manage its coherent usage.
}

%%%%%%%%%%%%%%%%%%%%%%%%%%
\subsection{IOMMUs}
\label{sec:iommu}
%%%%%%%%%%%%%%%%%%%%%%%%%%
Input-Output Memory Management Units (IOMMUs) are located between PCIe devices and the main memory.
Usually, they are implemented as part of the root complex.
Modern server systems feature one IOMMU per root port.
Similar to MMUs in the CPU, IOMMUs provide address translation and protection for memory regions that are made accessible to PCIe devices~\cite{intel2019vtd,amd2021iommu}.
Address virtualization allows to isolate or virtualize such devices.
Also, it allows 32-bit peripherals to use memory regions above \unit[4]{GB}.

The translation process of the IOMMU works very similar to the process in a CPU's MMU.
Modern IOMMUs map PCIe devices to IOMMU groups or domains.
The operating system, hypervisor, or VMM maintains a page table with all address mappings per group/domain.
The page table is organized in a tree structure.
Its depth depends on the width of the I/O virtual addresses (IOVAs) supported by the IOMMU.
For IOVAs referencing \unit[4]{KB} pages, the 12 least significant address bits (page offset) remain untranslated.
Accordingly, the 21/30 least significant bits of IOVAs pointing to \unit[2]{MB}/\unit[1]{GB} pages remain untranslated.

IOVAs are translated to physical addresses (PAs) by the IOMMU performing a page table walk.
Since this is quite time consuming, modern IOMMUs feature a translation look-aside buffer called IOTLB.
This cache is used to store translated IOVA$\to$PA mappings and is shared by all devices managed by the IOMMU.

%%%%%%%%%%%%%%%%%%%%%%%%%%%%%%%%%%%%%
\subsubsection{Attacks on IOMMUs}
\label{sec:iomm-attacks}
%%%%%%%%%%%%%%%%%%%%%%%%%%%%%%%%%%%%%
In the past, several attacks have been shown that circumvent the IOMMU to gain direct memory access or use the misconfiguration of the IOMMU to exploit device drivers through code injection or control-flow hijacking. However, the root cause always was a misconfigured IOMMU or a software vulnerability. We are not aware of any attacks that were made possible solely by the IOMMU hardware.

For example, a malicious peripheral can bypass the IOMMU by adding appropriate entries to the page table on startup before the IOMMU is activated by the BIOS~\cite{morgan2016bypassing,morgan2018iommuprotection}, or by exploiting PCIe address translation services (ATS), which allows a peripheral to mark any memory request as ``translated'' and bypass IOMMU translation and isolation~\cite{markettos2019thunderclap}.
Malicious devices may also exploit vulnerabilities in the kernel or device drivers.
IOMMU address translation only works on a page-granular level, so memory that was never intended to be shared might be allocated to a shared page, leaking secret data or enabling code injection attacks that can compromise the whole system~\cite{markettos2019thunderclap}.

%%%%%%%%%%%%%%%%%%%%%%%%%%%%%%%%%%%%%%%%%%%%%
\section{Identifying IOTLB Side-Channels}
\label{sec:reverse-engineering}
%%%%%%%%%%%%%%%%%%%%%%%%%%%%%%%%%%%%%%%%%%%%%

In this section, we demonstrate two fundamental techniques for implementing IOTLB side-channel attacks on these or similar systems.
We measure the latency difference between DMA accesses to addresses with cached and uncached translations in the IOMMU.
We also demonstrate a new algorithm for reliably finding IOTLB eviction sets with no prior assumptions about size or organization.
We have access to three different system setups that we will investigate throughout this work. 
\autoref{tab:setups} summarizes the key features of each. A detailed description of the setups is given next.

%%%%%%%%%%%%%%%%%%%%%%%%%%%%%%%%
\subsection{System Setup}
\label{app:setups}
%%%%%%%%%%%%%%%%%%%%%%%%%%%%%%%%

\begin{table}[t]
  \centering
  \caption{Overview of the system setups used in this work}
  \label{tab:setups}
  \begin{tabular}{p{.205\linewidth}p{.205\linewidth}p{.205\linewidth}p{.205\linewidth}}
    \toprule
    Name                        & \emph{a10l}          & \emph{a10v}          & \emph{s10v}          \\
    \midrule
    CPU                         & 2 Xeon Silver 4114   & 2 Xeon Platinum 8180 & 2 Xeon Platinum 8280 \\
    \#PCIe RP                   & 4 per CPU            & 4 per CPU            & 4 per CPU            \\
    \#IOMMUs                    & 4 per CPU            & 4 per CPU            & 4 per CPU            \\
    FPGA PAC                    & Arria 10             & Arria 10             & Stratix 10           \\
    OPAE ver.                   & 1.1.2-1              & 2020-01-01           & 2020-01-01           \\
    Bitstream ver.              & 1.2.3                & 1.1.3                & 2.0.3                \\
    Root/phys. access           & yes                  & no                   & no                   \\
    \bottomrule
  \end{tabular}
\end{table}

\revision{
For our experiments, we rely on three systems that are representative of modern cloud services featuring FPGA resources. 
The systems feature recent server-grade CPUs as well as FPGA extension cards based on Intel FPGAs. 
The FPGAs are managed by the Intel Acceleration Stack (IAS) which is designed to ease management of cloud deployments. 
The first system, \emph{a10l}, is a system we have physical and administrative access to. 
The other two systems \emph{a10v} and \emph{s10v} are cloud-like systems that are accessible through the Intel Labs (IL) Academic Compute Environment (ACE)\footnote{https://wiki.intel-research.net/}. 
We operate the two IL ACE systems with user privileges only. 
This is why we evaluate our eviction set finding algorithm on all three systems but rely solely on \emph{a10l} for the side- and covert channel experiments. 
More detailed information about the different systems is given in \autoref{tab:setups} and in the following paragraphs.
}

\emph{a10l}: As our local setup, we use a Dell PowerEdge R740 server
with two Intel Xeon Silver 4114 CPUs.
Each CPU reports 4 PCIe root bridges with one IOMMU per root port.
The system contains a Realtek PCIe ethernet network interface card (NIC).
It is assigned to a dedicated IOMMU group.
The NIC is passed-through to a virtual machine (VM) on the server.
An ethernet cable connects the NIC with one of the on-board NICs.
An NVIDIA Tesla T4 GPU 
is assigned to another dedicate IOMMU group that is managed by a different IOMMU than the NIC. 
Therefore, the NIC and the GPU do not share an IOTLB.
An Intel Programmable Acceleration Card (PAC) with Intel Arria 10 GX FPGA shares the IOTLB with the NIC or the T4, depending on the experiment, by connecting it to PCIe slots that are managed by the IOMMU also managing the NIC or the GPU respectively. 
All other PCIe devices like the on-board NICs, memory controllers, etc. are connected to different IOMMUs and therefore cannot interfere with our measurements.
The system has IAS 1.2 installed which contains OPAE version 1.1.2-1.
Running \texttt{fpgainfo} reports bitstream id \texttt{0x123000200000185} and bitstream version \texttt{1.2.3}.
We execute the GPU-accelerated database OmniSciDB\footnote{\url{https://docs.omnisci.com/overview/overview\#omniscidb}} in version 5.10., which is the latest version at the time of writing.
Additionally, CUDA version 11.4 and GPU driver version 470.57.02 are installed.
The database consists of one table filled with the Meta Kaggle data set\footnote{\url{https://www.kaggle.com/kaggle/meta-kaggle}}.
We have root access to this machine.

\emph{a10v}: The IL ACE contains servers with 
two Intel Xeon Platinum 8180 CPUs.
Each CPU reports 4 PCIe root bridges with IOMMU per root port.
Two PCIe PACs with Arria 10 GX FPGAs are managed by two separate IOMMUs. 
All other PCIe devices are managed by other IOMMUs.
The servers use IAS 1.1 and OPAE was installed on 01/01/2020 from the Git repository.
Running \texttt{fpgainfo} reports bitstream id \texttt{0x113000200000177} and bitstream version \texttt{1.1.3}.
We operate these machines with user privileges only.

\emph{s10v}: The IL ACE features servers with 
two Intel Xeon Platinum 8280 CPUs.
Each CPU reports 4 PCIe root bridges with one IOMMU per root port.
An Intel FPGA PAC D5005 is connected via PCIe. 
All other PCIe devices are managed by other IOMMUs than the one managing the PAC.
The servers use IAS 2.0 and OPAE was installed on 01/01/2020 from the Git repository.
Running \texttt{fpgainfo} reports bitstream version \texttt{2.0.3} and bitstream id \texttt{0x203000200000339}.
We operate these machines with user privileges only.

%%%%%%%%%%%%%%%%%%%%%%%%%%%%%%%%%%%%%%%%%%%%%%%%%%
\subsection{IOTLBs Cause Timing Behavior}
\label{sec:iotlb-timing-difference}
%%%%%%%%%%%%%%%%%%%%%%%%%%%%%%%%%%%%%%%%%%%%%%%%%%
During their PCIe performance benchmarking, Neugebauer et al. \cite{neugebauer2018pcieperformance} found that an IOTLB miss results in a latency increase of \unit[330]{ns}.
Since the FPGAs in our systems are clocked at \unit[200]{MHz}, the expected difference between fast and slow accesses is 66 clock cycles.
Peglow's \cite{peglow2020iommu} work matches our expectation.
With disabled IOMMU, the memory read latency for any address in main memory is distributed around 160 and 185 cycles.
When the system is configured to use the IOMMU, this distribution shifts to 225 and 270 cycles for addresses that are accessed for the first time.
Access times for subsequent accesses are distributed similarly to access times measured without IOMMU.
Thus the measurable latency difference between accesses to addresses where the translation is present in or absent from the IOTLB lies between 65 and 85 clock cycles.
We reproduced all values for the \emph{a10l} system. On the IL ACE systems \revision{\emph{a10v} and s10v}, the latency difference between first accesses and subsequent accesses lies in the expected range.
However, we cannot disable the IOMMU on the IL ACE systems to check whether the latency difference disappears.

%%%%%%%%%%%%%%%%%%%%%%%%%%%%%%%%%%%%%%%%%%%%%%%%%%
\subsection{Tools for Testing IOMMU Behavior}
\label{sec:tools-iommu}
%%%%%%%%%%%%%%%%%%%%%%%%%%%%%%%%%%%%%%%%%%%%%%%%%%
The IOMMU translates addresses for peripherals.
Therefore, the CPU alone can only interact with the IOMMU in limited ways; we have to rely on a peripheral device to perform the experiments.
For this purpose we used the PCIe PACs with DMA capabilities.
We implement a hardware function for the FPGA that is programmable from software to capture the required measurements.

%%%%%%%%%%%%%%%%%%%%%%%%%%%%%%%%%%%%%%%%%%%%%%%%%%%%%%%%%
\subsubsection{IOTLB Control from the CPU}
\label{sec:iommu-control-cpu}
%%%%%%%%%%%%%%%%%%%%%%%%%%%%%%%%%%%%%%%%%%%%%%%%%%%%%%%%%
To assist with these experiments, we also develop a kernel module that enables a program on the CPU to flush all entries from the IOTLB of a given IOMMU.
When loaded, the kernel module uses a variety of functions and structures from the Linux kernel source, including those found in \texttt{<linux/pci.h>}, \texttt{<linux/iommu.h>}, and \texttt{<linux/dmar.h>} to find a PCIe device structure based on its vendor and device IDs, and from there find the device structure corresponding to the IOMMU that manages that PCIe device.
That IOMMU device structure already contains a pointer to a function for flushing the IOMMU, so that function merely needs to be called.
The kernel module uses a character file and ioctl as an interface by which user programs can call for the kernel module to flush the IOMMU.
However, it takes root access to load a kernel module, since the module must read and write kernel memory.
Therefore, we only tested \autoref{alg:evicts} with the optional flush on our local system \emph{a10l}.

%%%%%%%%%%%%%%%%%%%%%%%%%%%%%%%%%%%%%
\subsubsection{Hardware Design}
\label{sec:hardware-design}
%%%%%%%%%%%%%%%%%%%%%%%%%%%%%%%%%%%%%
Our \texttt{iotlb\_pnp} hardware module is designed against the Intel Acceleration Stack as would be the case in a cloud environment.
The module is capable of performing memory accesses and timing the access latency.
\texttt{iotlb\_pnp} can be programmed with up to 7 instructions.
Currently, the design supports 5 instructions: \texttt{evset\_prime}, \texttt{evset\_probe}, \texttt{target\_prime}, \texttt{target\_probe}, and \texttt{wait}.
Configuration and programming of the hardware module is performed via MMIO through OPAE.
The prime instructions make the hardware module access a configured address (target) or set of addresses (eviction set).
Probe instructions behave in the same way as the prime instructions but additionally count clock cycles.
When probing an eviction set, the module can be configured to either measure the overall execution time of the instruction or time each memory access individually.
The eviction sets used during priming and probing can be configured independent from each other, as is the case for the target instructions.
The wait instruction simply makes the hardware module do nothing for a configured number of clock cycles.

%%%%%%%%%%%%%%%%%%%%%%%%%%%%%%%%%%%%%%%%%%%%%%
\subsubsection{Software}
\label{sec:software}
%%%%%%%%%%%%%%%%%%%%%%%%%%%%%%%%%%%%%%%%%%%%%%
The software counterpart to the hardware module uses the OPAE C library to interact with the hardware design on the FPGA. 
This library allows us to control and observe the operation of the hardware module with memory-mapped I/O (MMIO) as well as --- crucially for the work that this module must do --- allocate shared pages of the system's main memory that the FPGA as well as the CPU can read and write.

%%%%%%%%%%%%%%%%%%%%%%%%%%%%%%%%%%%%%%%%%%%%%%%%%%%%%%%%%
\revision{
\subsection{Threat Models}
\label{sec:threatmodel}
}
%%%%%%%%%%%%%%%%%%%%%%%%%%%%%%%%%%%%%%%%%%%%%%%%%%%%%%%%%
 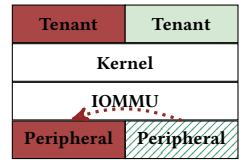
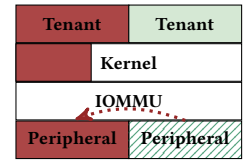
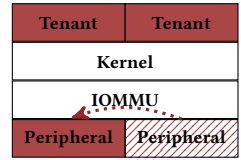
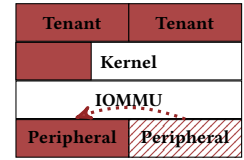
\begin{figure}[t]
   \centering
   \begin{subfigure}[t]{.45\linewidth}
     \centering
     \begin{tikzpicture}[
node distance=1.2cm,
every node/.style={draw=black},
alignc/.style={align=center},
alignl/.style={align=left},
alignr/.style={align=right},
small dim/.style={minimum height=.5cm, minimum width=1.5cm, text width=1.25cm},
wide dim/.style={minimum height=.5cm, minimum width=3cm, text width=2.75cm},
attacker/.style={fill=niceredfill},
victim/.style={fill=nicegreenfill},
victim partial/.style={pattern={north east lines},pattern color=nicegreen},
red/.style={draw=niceredborder, fill=niceredfill, text=white, very thick},
]
\definecolor{nicegreen}{rgb}{0.335938, 0.640625, 0.496094},
\tikzset{every node} = [font=\footnotesize\bfseries]

\node[wide dim, alignc] (kernel) {Kernel};
\node[wide dim, alignc, anchor=north]  (iommu) at (kernel.south) {IOMMU};

\node[small dim, alignc, anchor = south west, attacker] (user-a) at (kernel.north west) {Tenant};
\node[small dim, alignc, anchor = south east, victim] (user-v) at (kernel.north east) {Tenant};

\node[small dim, alignc, anchor = north west, attacker] (dev-a) at (iommu.south west) {Peripheral};
\node[small dim, alignc, anchor = north east, victim partial] (dev-v) at (iommu.south east) {Peripheral};
\path[draw, red, dotted, very thick] (dev-v.north)  edge[->, bend right=20] (dev-a.north);
\end{tikzpicture}
     \caption{\revision{Model \emph{1u}. Side-channel attacker with \emph{user} privilege.}}
   \end{subfigure}
   \hfill
   \begin{subfigure}[t]{.45\linewidth}
     \centering
     \begin{tikzpicture}[
node distance=1.2cm,
every node/.style={draw=black},
alignc/.style={align=center},
alignl/.style={align=left},
alignr/.style={align=right},
small dim/.style={minimum height=.5cm, minimum width=1.5cm, text width=1.25cm},
xs dim/.style={minimum height=.5cm, minimum width=1cm, text width=.5cm},
wide dim/.style={minimum height=.5cm, minimum width=3cm, text width=2.75cm},
attacker/.style={fill=niceredfill},
victim/.style={fill=nicegreenfill},
victim partial/.style={pattern={north east lines},pattern color=nicegreen},
red/.style={draw=niceredborder, fill=niceredfill, text=white, very thick},
]
\definecolor{nicegreen}{rgb}{0.335938, 0.640625, 0.496094},
\tikzset{every node} = [font=\footnotesize\bfseries]

\node[wide dim, alignc] (kernel) {Kernel};
\node[wide dim, alignc, anchor=north]  (iommu) at (kernel.south) {IOMMU};

\node[small dim, alignc, anchor = south west, attacker] (user-a) at (kernel.north west) {Tenant};
\node[small dim, alignc, anchor = south east, victim] (user-v) at (kernel.north east) {Tenant};

\node[small dim, alignc, anchor = north west, attacker] (dev-a) at (iommu.south west) {Peripheral};
\node[small dim, alignc, anchor = north east, victim partial] (dev-v) at (iommu.south east) {Peripheral};

\node[xs dim, alignc, anchor = north west, attacker] (module) at (kernel.north west) {};

\path[draw, red, dotted, very thick] (dev-v.north)  edge[->, bend right=20] (dev-a.north);
\end{tikzpicture}
     \caption{\revision{Model \emph{1k}. Side-channel attacker with \emph{kernel} module.}}
   \end{subfigure}
   \newline\newline
   \begin{subfigure}[t]{.45\linewidth}
     \centering
     \begin{tikzpicture}[
node distance=1.2cm,
every node/.style={draw=black},
alignc/.style={align=center},
alignl/.style={align=left},
alignr/.style={align=right},
small dim/.style={minimum height=.5cm, minimum width=1.5cm, text width=1.25cm},
wide dim/.style={minimum height=.5cm, minimum width=3cm, text width=2.75cm},
attacker/.style={fill=niceredfill},
attacker partial/.style={pattern={north east lines},pattern color=nicered},
victim/.style={fill=nicegreenfill},
victim partial/.style={pattern={north east lines},pattern color=nicegreen},
red/.style={draw=niceredborder, fill=niceredfill, text=white, very thick},
]
\definecolor{nicegreen}{rgb}{0.335938, 0.640625, 0.496094},
\definecolor{nicered}{rgb}{0.667969, 0.273438, 0.273438},
\tikzset{every node} = [font=\footnotesize\bfseries]

\node[wide dim, alignc] (Kernel) {Kernel};
\node[wide dim, alignc, anchor=north]  (iommu) at (kernel.south) {IOMMU};

\node[small dim, alignc, anchor = south west, attacker] (user-a) at (kernel.north west) {Tenant};
\node[small dim, alignc, anchor = south east, attacker] (user-v) at (kernel.north east) {Tenant};

\node[small dim, alignc, anchor = north west, attacker] (dev-a) at (iommu.south west) {Peripheral};
\node[small dim, alignc, anchor = north east, attacker partial] (dev-v) at (iommu.south east) {Peripheral};
\path[draw, red, dotted, very thick] (dev-v.north)  edge[->, bend right=20] (dev-a.north);
\end{tikzpicture}
     \caption{\revision{Model \emph{2u}. Covert channel with \emph{user} privilege.}}
   \end{subfigure}
   \hfill
   \begin{subfigure}[t]{.45\linewidth}
     \centering
     \begin{tikzpicture}[
node distance=1.2cm,
every node/.style={draw=black},
alignc/.style={align=center},
alignl/.style={align=left},
alignr/.style={align=right},
small dim/.style={minimum height=.5cm, minimum width=1.5cm, text width=1.25cm},
wide dim/.style={minimum height=.5cm, minimum width=3cm, text width=2.75cm},
xs dim/.style={minimum height=.5cm, minimum width=1cm, text width=.5cm},
attacker/.style={fill=niceredfill},
attacker partial/.style={pattern={north east lines},pattern color=nicered},
victim/.style={fill=nicegreenfill},
victim partial/.style={pattern={north east lines},pattern color=nicegreen},
red/.style={draw=niceredborder, fill=niceredfill, text=white, very thick},
]
\definecolor{nicegreen}{rgb}{0.335938, 0.640625, 0.496094},
\definecolor{nicered}{rgb}{0.667969, 0.273438, 0.273438},
\tikzset{every node} = [font=\footnotesize\bfseries]

\node[wide dim, alignc] (kernel) {Kernel};
\node[wide dim, alignc, anchor=north]  (iommu) at (kernel.south) {IOMMU};

\node[small dim, alignc, anchor = south west, attacker] (user-a) at (kernel.north west) {Tenant};
\node[small dim, alignc, anchor = south east, attacker] (user-v) at (kernel.north east) {Tenant};

\node[small dim, alignc, anchor = north west, attacker] (dev-a) at (iommu.south west) {Peripheral};
\node[small dim, alignc, anchor = north east, attacker partial] (dev-v) at (iommu.south east) {Peripheral};
\node[xs dim, alignc, anchor = north west, attacker] (module) at (kernel.north west) {};
\path[draw, red, dotted, very thick] (dev-v.north)  edge[->, bend right=20] (dev-a.north);
\end{tikzpicture}
     \caption{\revision{Model \emph{2k}. Covert channel with \emph{kernel} module.}}
   \end{subfigure}
   \caption{\revision{Comparison of threat models. Dark red fills indicate functional units controlled by a malicious actor, and light green fills indicate functional units controlled by a victim. Diagonal lines indicate functional units that are only under coarse or indirect control, e.g., a simple network interface card or an accelerator that assists with certain applications but is not directly programmable. The dashed arrows indicate the flow of data through the channel.}}
   \label{fig:threat_1u}
 \end{figure}

\revision{
We consider two general threat models with two variants each, as illustrated in \autoref{fig:threat_1u}.
All four threat models include a malicious actor that can program and control a fast and programmable PCIe device (referred to in this section as the monitoring device) with direct memory access (such as an FPGA or GPU) and an IOMMU providing address translation services for that device. 
Each model also includes a second peripheral (referred to in this section as the sending device) which also uses the same IOMMU for DMA address translation but does not need to be fast or directly programmable as part of the threat model. 
The monitoring device must be capable of timing memory accesses and reliably differentiate IOTLB hits from misses.
The attacker must further be able to program the monitoring device directly to find eviction sets and execute Prime+Probes.
The sending device only needs to have memory access patterns that can be triggered by a user, either by direct control, or triggerable through an application or system interface.
}

\revision{
Models \emph{1k} and \emph{1u} are adversarial threat models for a \emph{side-channel attack}, where a malicious user in control of the monitoring device exploits IOTLB contention to gain secret information from another user's application that triggers memory accesses in the sending device. 
Models \emph{2k} and \emph{2u} outline the requirements for a \emph{covert channel} with cooperative sending and monitoring devices, where colluding malicious users in control of applications in separate security domains uses the IOTLB to transmit data covertly across the two devices.
Models ending in \emph{k} include kernel access alongside the monitoring device, and models ending in \emph{u} do not. 
Kernel access is necessary to implement an IOTLB flush through a custom kernel module as outlined in \autoref{sec:tools-iommu}.
In \autoref{sec:evset-construction} we show how fine-grained flushing control allows for more reliable eviction set construction.
However, eviction set construction and Prime+Probe-based IOTLB side-channel attacks are still possible without flushing capabilities.
}

\revision{
Whereas some side-channel attacks can be carried out with JavaScript from a web browser against a personal computer, we consider cloud environments as the primary site of IOTLB attacks, since the attacker must already have control of a peripheral. 
Renting a single GPU or FPGA in a cloud environment is easy; the primary logistical challenge of setting up a practical IOTLB side-channel or covert channel is IOMMU co-location -- that is, ensuring that the monitoring device shares an IOMMU (and IOTLB) with the sending device.
However, research into similar problems, like co-locating cloud instances for cache attacks, has yielded strategies for co-location that can be adapted to the IOTLB channel. 
\.{I}nci et al.~\cite{inci2016co} demonstrated two reliable co-location techniques for last-level caches that rely only on basic cache contention and so could be adapted to the IOTLB relatively easily.
In a cooperative (covert channel) scenario, the sender instance sends a predetermined signal and the receiving instance searches the channel for a signal and attempts to match it with the agreed-upon signal.
In an adversarial (attack) scenario, the attacker first chooses a target program and profiles it locally to learn to identify the traces it leaves. Then the attacker searches for such traces. For cache profiling, co-location is not necessary; the target program can be profiled within a single instance. In the case of IOTLB profiling, covert channel co-location may be used to first co-locate the cloud instance controlling the monitoring peripheral with another cloud instance that runs the target program which relies on a sending peripheral.
}

%%%%%%%%%%%%%%%%%%%%%%%%%%%%%%%%%%%%%%%%%%%%%%%%%%%%%%%%%%%%%%%%%%%%%%%%%%
\section{Constructing Eviction Sets}
\label{sec:evset-construction}
%%%%%%%%%%%%%%%%%%%%%%%%%%%%%%%%%%%%%%%%%%%%%%%%%%%%%%%%%%%%%%%%%%%%%%%%%%

Initially, we hypothesized that the IOTLB would be organized like the CPU TLBs reverse-engineered in \cite{gras2018translation}, with $2^s$ sets where $s$ is an integer, some small number of ways per set, and a set mapping algorithm wherein the lowest $s$ bits of the page address select the set number or some other combination of various bits of the page address forms the set number that the page is associated with.
\revision{
However, as we describe in \autoref{sec:initial_hypothesis} in the appendix, this turned out to be false. So we set out to construct eviction sets for IOTLBs with an unknown architecture.
}

%%%%%%%%%%%%%%%%%%%%%%%%%%%%%%%%%%%%%%%%%%%%%%%%%%%%%%%%%%%%%%%%%%%%
\subsection{A New Approach to Eviction Set Construction}
%%%%%%%%%%%%%%%%%%%%%%%%%%%%%%%%%%%%%%%%%%%%%%%%%%%%%%%%%%%%%%%%%%%%

\noindent
\begin{table}[t]
  \caption{Notation used in algorithms}
  \label{tab:algonotation}
  \centering
  \begin{tabular}{ll}
    \toprule
    Symbol & Meaning \\
    \midrule
    $A \gets B$     & $A$ gets the value of $B$ \\
    $A \gets_\in B$ & $A$ chosen randomly from $B$ \\
    $A \gets_+ B$   & $B$ added to the set $A$ \\
    $A \gets_- B$   & $B$ removed from the set $A$ \\
    $A \gets_/ B$   & Elements in $B$ removed from $A$ \\
    \bottomrule
  \end{tabular}
\end{table}

%\SetKwProg{Fn}{Function}{}{}
%\SetKwInOut{Input}{input}
%\SetKwInOut{Output}{output}
%\SetKwFunction{evicts}{evicts}
\begin{algorithm}[t]
    \footnotesize
    \DontPrintSemicolon
    \Fn{\evicts{target, evset}}{
        \Input{\emph{target} -- address to be evicted\\\emph{evset} -- eviction set used for eviction attempt}
        \Output{True, if 100 eviction attempts are successful\\False, otherwise}
        count $\gets$ 0 \tcp*[h]{\# of contentions}\;
        \For{$0 \leq i < 100$}{
            flush IOTLB \tcp*[h]{optional}\;
            \texttt{target\_prime}()\;
            \texttt{evset\_prime}()\;
            time $\gets$ \texttt{target\_probe}()\;
            \If{time > threshold}{
                count $\gets$ count + 1\;
            }
        }
        \Return count == 100\;
    }
    \caption{\bfseries The algorithm tests whether a given eviction set evicts a given target address from the IOTLB. The target\_prime and evset\_prime function calls have the FPGA access the respective set of addresses. The function call target\_probe has the FPGA time the access time to the target address.}
    \label{alg:evicts}
\end{algorithm}

\revision{
We developed a novel and platform-independent algorithm for finding eviction sets for any TLB or cache where the timing difference between a present entry and an evicted entry is known and measurable.
Our approach is inspired by the baseline reduction algorithm in \cite{vila2019theory}, which only reduces an already existing eviction set to its minimum necessary size, and the grow-split eviction set construction approach of Algorithm 1 in \cite{liu2015last}.
}

\revision{
Like \cite{liu2015last}, our algorithm constructs eviction sets from a large pool of addresses by gathering candidates for an eviction set and then systematically discarding unnecessary ones; addresses not present in candidate eviction sets are used as test targets. 
The grow-split algorithm in \cite{liu2015last} is specifically designed for a partitioned cache: it first constructs an eviction set for the entire cache, and then splits it into separate sets for each of the partitions.
Our grow-reduce algorithm makes no assumption about cache organization, and uses a more generalized approach of building one eviction set at a time by adding addresses until evictions are reliable and then testing which addresses can be discarded without losing reliability.
It aims to create an exhaustive set of eviction sets by searching the entire address pool; redundant sets are avoided by ensuring that potential test targets are not already reliably evicted by another set.
}

\subsubsection{\revision{Grow-Reduce Algorithm}}
~~~~

%\SetKwFunction{constructEvset}{constructEvset}
\begin{algorithm}[t]
    \footnotesize
    \DontPrintSemicolon
    \Fn{\constructEvset{target, pool}}{
        \Input{\emph{target} -- target address to be evicted\\\emph{pool} -- address pool}
        \Output{\emph{evset} -- an eviction set for \emph{target}}
        evset $\gets \emptyset$\;
        count $\gets 0$ \tcp*[h]{\# of contentions}\;
        \tcp*[l]{Grow}
        \While{count < 50 and $|$pool$|$ > 0}{
            page $\gets_\in$ pool;
            evset $\gets_+$ page;
            pool $\gets_-$ page\;
            \If{$\mathsf{evicts}$(target, evset)}{
                count $\gets$ count + 1\;
            }
        }
        \tcp*[l]{Reduce}
        \ForEach{page in evset}{
            evset $\gets_-$ page\;
            \If{not $\mathsf{evicts}$(target, evset)}{
                evset $\gets_+$ page\;
            }
        }
        \Return evset\;
    }
    \caption{\bfseries The algorithm constructs an IOTLB eviction set for a given target address. The addresses for the eviction set are chosen from the given address pool.}
    \label{alg:constructEvset}
\end{algorithm}

%\SetKwFunction{evsetFinding}{evsetFinding}
\begin{algorithm}[t]
    \footnotesize
    \DontPrintSemicolon
    \Fn{\evsetFinding{poolSize}}{
        \Input{\emph{poolSize} -- number of addresses to be allocated}
        \Output{\emph{evsets} -- Eviction sets for the IOTLB}
        pool $\gets \mathsf{alloc}$(poolSize)\;
        targets $\gets \emptyset$\;
        evsets $\gets \emptyset$\;
        \While{poolSize > 0}{
            target $\gets_\in$ pool \tcp*[h]{Random page as target}\;
            pool $\gets_-$ target\;
            \If{evsets do not evict target}{
                targets $\gets_+$ target\;
                evsets $\gets_+ \mathsf{constructEvset}$(target, pool)\;
                pool $\gets_/$ evsets\;
            }
            poolSize $\gets$ size(pool);
        }
        \Return evsets
    }
    \caption{\bfseries This algorithm constructs as many eviction sets as needed to evict any target address from the IOTLB. The algorithm takes an integer as input that indicated the size of the address pool that is used to construct the eviction sets. A pool size of 4096 was used for the tests in this paper.}
    \label{alg:evsetfinding}
\end{algorithm}

The most basic function in our algorithm tests whether or not a hypothetical eviction set evicts a given target address (see \autoref{alg:evicts}).
The software uses the hardware module described previously to perform a prime and probe test.
First, the FPGA accesses the target followed by an access to each address in the eviction set.
Then the target is accessed again and the access latency is measured.
We define that an eviction set evicts a target if the latency of the second access to the target is above a certain threshold.
We choose the threshold in the middle of the observed latency gap between fast and slow accesses observed on the different systems.
Before each prime and probe test, we optionally cleared the IOTLB.

The construction of an eviction set for a fixed target address is given in \autoref{alg:constructEvset}.
It takes a target address and a pool of addresses as inputs.
The eviction set is initialized as an empty set.
During the "grow" step random addresses are chosen from the address pool and added to the eviction set until the eviction set contains enough addresses to evict the target.
Obviously, the eviction set may contain unnecessary addresses at this point.
This is why a reduction step follows where each address is tested for its necessity.
If an address is not needed, it is removed from the eviction set and put back in the address pool.

At the highest level, our algorithm shown in \autoref{alg:evsetfinding} automatically constructs as many eviction sets as it can find.
The program first allocates a pool of memory pages.
\revision{
For our experiments we used a pool size of 4096 addresses.
}
The algorithm manages two sets: The \emph{targets} set is used to store the different target addresses used during eviction set construction.
The \emph{evsets} set stores all eviction sets constructed by the algorithm.
After this initialization step, the algorithm picks a random target address from the pool and removes it from the pool.
If \emph{evsets} does not contain an eviction set for the target address yet, a new eviction set is constructed.
The target address and the new eviction set are added to their corresponding sets.
All addresses in the newly constructed eviction set are then removed from the pool.
This procedure is repeated until the pool does not contain any addresses anymore.

%%%%%%%%%%%%%%%%%%%%%%%%%%%%%%%%%%%%%%%%%%%
\subsubsection{Evaluation of New Eviction Set Algorithm}
\label{sec:experimental-results}
%%%%%%%%%%%%%%%%%%%%%%%%%%%%%%%%%%%%%%%%%%%
We found that the optional flushing of the IOTLB has an impact on the size and reliability of IOTLB eviction sets. 
\footnote{\revision{Flushing the IOTLB requires kernel access; see threat models \emph{1k} and \emph{2k} in \autoref{sec:threatmodel}. For this reason, \autoref{tab:algstats} contains data only from experiments on the \emph{a10l} system.}}
\revision{
The major differences are laid out in \autoref{tab:algstats}, which enumerates general performance metrics of eviction sets constructed with our grow-reduce algorithm and \cite{liu2015last}'s grow-split algorithm both with and without flushing.
}
\revision{
Enabling IOTLB flushes before the Prime+Probe step will make both algorithms return a single eviction set containing 118 addresses.
The success rate of such eviction sets is 100\%  in every case we observed.
}

\revision{
Without IOTLB flushes, neither algorithm produces such consistently sized or reliable eviction sets.
This is likely due to a replacement policy that we were unable to deduce.
In this scenario we can better see the advantage of our grow-reduce algorithm.
It produces eviction sets that are both smaller and twice as reliable than those produced by the grow-split algorithm.
}

\begin{table*}[t]
  \centering
  \caption{\revision{Comparison of eviction set finding algorithms on the IOTLB of the \emph{a10l} test system. All tests were conducted on the \emph{a10l} system using pools of 4096 addresses, and repeated 40 times. Eviction set orders were randomized between prime and probe steps during testing.}}
  \label{tab:algstats}
  \begin{tabular}  { l c l r r r r }
  \toprule
  Flush & &
    Algorithm & Number of sets & Set size & Useful sets per target & Average best eviction rate \\
  \midrule
  \multirow{2}{*}{enabled}
    & \begin{imageonly}\ldelim\{{2}{2pt}\end{imageonly} 
    & Grow-Reduce (this work)         &  1.00 & 118.00 & 1.00 & 100.00\,\% \\ 
  & & Grow-Split (\cite{liu2015last}) &  1.00 & 118.00 & 1.00 & 100.00\,\% \\
  \multirow{2}{*}{disabled}
    & \begin{imageonly}\ldelim\{{2}{2pt}\end{imageonly} 
    & Grow-Reduce (this work)         & 32.08 & 110.05 & 0.98 &  82.23\,\% \\
  & & Grow-Split (\cite{liu2015last}) & 10.70 & 50.69  & 0.98 &  28.00\,\% \\
  \bottomrule
  \end{tabular}
\end{table*}

\begin{figure*}[t]
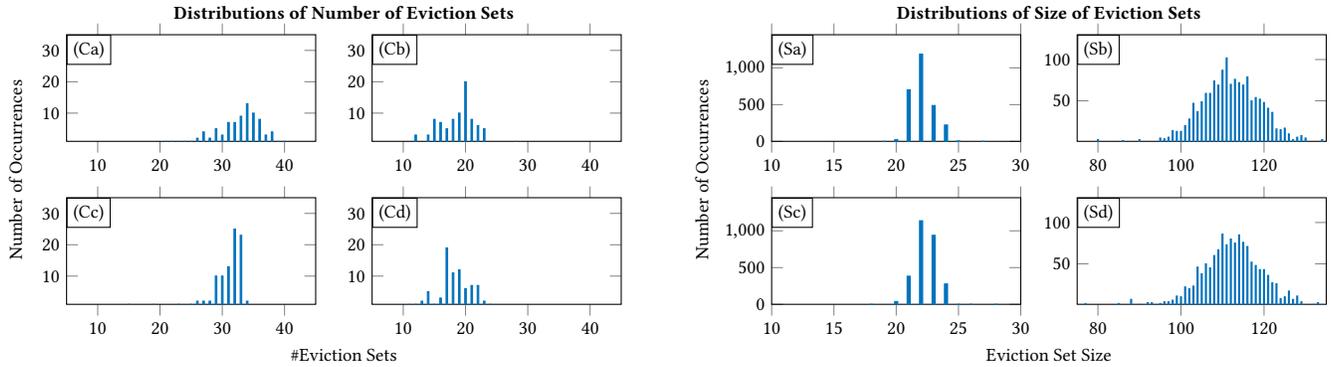

    \centering
    \pgfplotsset{
    BlueBars/.style={
        draw=plotblue,
        fill=plotblue,
        bar width=0.25
    },
}
\pgfplotsset{select coords between index/.style 2 args={
    x filter/.code={
        \ifnum\coordindex<#1\def\pgfmathresult{}\fi
        \ifnum\coordindex>#2\def\pgfmathresult{}\fi
    }
}}

\pgfplotstableread[col sep=comma,]{figures/evsets/data/evset_count_r0_s0_s10.csv}\crzsz
\pgfplotstableread[col sep=comma,]{figures/evsets/data/evset_count_r1_s0_s10.csv}\crosz
\pgfplotstableread[col sep=comma,]{figures/evsets/data/evset_count_r0_s1_s10.csv}\crzso
\pgfplotstableread[col sep=comma,]{figures/evsets/data/evset_count_r1_s1_s10.csv}\croso

\pgfplotstableread[col sep=comma,]{figures/evsets/data/evset_size_r0_s0_s10.csv}\srzsz
\pgfplotstableread[col sep=comma,]{figures/evsets/data/evset_size_r1_s0_s10.csv}\srosz
\pgfplotstableread[col sep=comma,]{figures/evsets/data/evset_size_r0_s1_s10.csv}\srzso
\pgfplotstableread[col sep=comma,]{figures/evsets/data/evset_size_r1_s1_s10.csv}\sroso

\begin{tikzpicture}
\tikzset{every node} = [font=\footnotesize]
\input{figures/evsets/evset_count_s10v}
\input{figures/evsets/evset_size_s10v}
\end{tikzpicture}
    \caption{Number of eviction sets and the size of each constructed set needed to evict any target IOVA after running \autoref{alg:evsetfinding} for 100 times each. During eviction set construction, randomization of the eviction set was turned off for measurements (a) and (c) and turned on for (b) and (d). For measurements (c) and (d), the algorithm waited \unit[100]{ns} between each eviction test. For measurements (a) and (b) this was not the case.
If the order of accesses during the \lstinline{evset_prime()} is static throughout one run of \autoref{alg:evsetfinding}, the resulting eviction sets contain 20 to 25 addresses each.
The average success rate is slightly below the average success rate of eviction sets constructed with randomized access order during \lstinline{evset_prime()}.
In turn, randomizing the access order yields on average slightly less but bigger sets.
The success rate of these sets, with or without randomized access order, evict a target with probabilities above 90\%.
    }
    \label{fig:evset_stats_s10v}
\end{figure*}

\revision{
\autoref{fig:evset_stats_s10v} visualizes in detail the results of further experimentation with small implementation tweaks in our algorithm.
In these experiments we found that the size and number of eviction sets constructed were very similar on all tested systems, \emph{a10l}, \emph{a10v}, and \emph{s10v}.
We thus conclude that the IOTLB architecture on all tested systems is very similar in terms of IOTLB size, organization and replacement policy.
}

%%%%%%%%%%%%%%%%%%%%%%%%%%%%%%%%%%%%%%%%%%%%%%%%%%%%
\section{Analysis of Side-Channel Leakages}
\label{sec:fingerprinting-attack}
%%%%%%%%%%%%%%%%%%%%%%%%%%%%%%%%%%%%%%%%%%%%%%%%%%%%
We now use the constructed eviction sets to further investigate the amount of leakage from PCIe devices observable in the IOMMU. Though we use the FPGA for channel monitoring outside of a virtualized environment for simplicity's sake, this channel still poses a threat from one virtual environment to another or from a virtual environment to hypervisor. Major cloud platforms like AWS and Alibaba Cloud now allow users to rent direct access to FPGAs with DMA capabilities, meaning that malicious tenants could easily run hardware designs that monitor the IOMMU side-channel without root privileges. Any other PCIe devices that are co-located on the IOMMU with a malicious FPGA and using translated DMA (most modern devices use DMA, and virtualized DMA always requires translation if the IOMMU is shared) are sources of leakage and therefore potential attack targets. 
We focus our analysis on an in-memory SQL database accelerated by a graphics card.

%%%%%%%%%%%%%%%%%%%%%%%%%%%%%%%%%%%%%%%%%%%%%%%%%%%%%%%%%%%%%%%%%%%%%%%%%%%%%%%%%%%%%%%
\subsection{GPU-Accelerated SQL Database Leakage}
\label{sec:sqltraces}
%%%%%%%%%%%%%%%%%%%%%%%%%%%%%%%%%%%%%%%%%%%%%%%%%%%%%%%%%%%%%%%%%%%%%%%%%%%%%%%%%%%%%%%
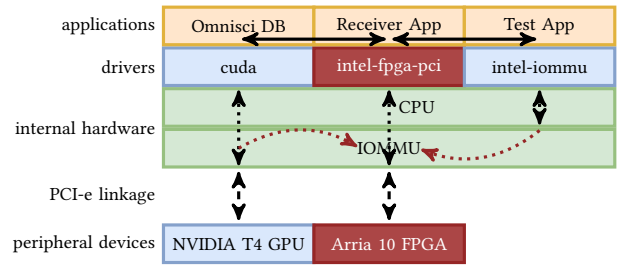
\begin{figure}[t]
    \centering
    \begin{tikzpicture}[
    node distance=1.2cm,
    alignc/.style={align=center},
    alignl/.style={align=left},
    alignr/.style={align=right},
    small dim/.style={minimum height=.5cm, minimum width=2cm, text width=1.75cm},
    wide dim/.style={minimum height=.5cm, minimum width=4cm, text width=3.75cm},
    medium wide dim/.style={minimum height=.5cm, minimum width=6cm, text width=5.75cm},
    very wide dim/.style={minimum height=.5cm, minimum width=8cm, text width=7.75cm},
    green/.style={draw=nicegreenborder, fill=nicegreenfill, very thick},
    blue/.style={draw=niceblueborder, fill=nicebluefill, very thick},
    red/.style={draw=niceredborder, fill=niceredfill, text=white, very thick},
    orange/.style={draw=niceorangeborder, fill=niceorangefill, very thick},
    yellow/.style={draw=niceyellowborder, fill=niceyellowfill, very thick},
]

\tikzset{every node} = [font=\footnotesize]

\node[medium wide dim, alignc, green] (cpu) {\quad\quad\quad CPU};
\node[medium wide dim, alignc, green, anchor=north]      (iommu) at (cpu.south) {IOMMU};

\node[small dim, alignc, blue, anchor=south west] (gpu-driver) at (cpu.north west) {cuda};
\node[small dim, alignc, red, anchor=south ]      (fpga-driver) at (cpu.north) {intel-fpga-pci};
\node[small dim, alignc, blue, anchor=south east] (iommu-driver) at (cpu.north east) {intel-iommu};

\node[small dim, alignc, orange, anchor=south] (sender0) at (gpu-driver.north) {Omnisci DB};
\node[small dim, alignc, orange, anchor=south] (receiver) at (fpga-driver.north) {Receiver App};
\node[small dim, alignc, orange, anchor=south] (sender1) at (iommu-driver.north) {Test App};

\node[small dim, alignc, blue, below=1cm of iommu.south west, anchor=west] (gpu)  {NVIDIA T4 GPU};
\node[small dim, alignc, red, below=1cm of iommu.south, anchor=center]  (fpga) {Arria 10 FPGA};

\node[text width=2cm, alignr, anchor=east] at (sender0.west) {applications};
\node[text width=2cm, alignr, anchor=east] at (gpu-driver.west) {drivers};
\node[text width=2cm, alignr, anchor=east] at (cpu.south west) {internal hardware};
\path (iommu.west) -- node [midway, text width=2cm, anchor=east, alignr] {PCI-e linkage} (gpu.west);
\node[text width=2cm, alignr, anchor=east] at (gpu.west) {peripheral devices};

\path[<->, draw, dashed, very thick] (gpu.north) -- (gpu.north |- iommu.south);
\path[<->, draw, dashed, very thick] (fpga.north) -- (fpga.north |- iommu.south);
\path[<->, draw, dotted, very thick] (gpu-driver.south) -- (gpu.north |- iommu.south);
\path[<->, draw, dotted, very thick] (fpga-driver.south) -- (fpga.north |- iommu.south);
\path[<->, draw, dotted, very thick] (iommu-driver.south) -- (iommu-driver.south |- iommu.north);
\path[draw, red, dotted, very thick] ([yshift=0pt]gpu.north |- iommu.center) edge[->, bend left=30] ([xshift=-11pt]fpga.north |- iommu.center);
\path[draw, red, dotted, very thick] ([yshift=-1pt]iommu-driver.south |- iommu.north) edge[->, bend left=30] ([xshift=11pt]iommu.center);
\path[<->, draw, very thick] ([yshift=-5pt]sender0.center) -- ([yshift=-5pt]receiver.center);
\path[<->, draw, very thick] ([yshift=-5pt]sender1.center) -- ([yshift=-5pt]receiver.center);
\end{tikzpicture}
    \caption{Stack diagram of the CPU to peripheral and peripheral to peripheral covert channel and side-channel tests.}
    \label{fig:sql_test_stack}
\end{figure}

We now inspect the amount of IOTLB leakage observable from the FPGA when it is co-located with a GPU that runs an SQL database.
For our tests, we co-locate the FPGA with an NVIDIA Tesla T4 GPU that runs the OmniSci SQL server on it.
\revision{We wish to understand the data leakage patterns of the GPU-accelerated database application, so for these experiments we consider threat model \emph{1k}, where the attacker has the most precise control over the channel.}
\autoref{fig:sql_test_stack} shows a stack diagram of the setup \revision{on our \emph{a10l} platform}.
The test application interacts with our hardware module on the FPGA to construct, prime and probe an eviction set for the IOTLB.
Additionally, the application can issue SQL queries to the database which computes the result on the GPU.

After constructing an eviction set for the IOTLB, the test app primes the IOTLB.
During the waiting phase, the app runs an SQL query on the GPU.
The tested queries differ (significantly) in the size of the returned results.
After the SQL result is returned to the test application, the FPGA probes the IOTLB and reports the access latency back to the application.
\begin{figure}[t]
    \centering
    \pgfplotsset{
    BlueBars/.style={
        draw=plotblue,
        fill=plotblue,
        bar width=0.25
    },
}
\pgfplotsset{select coords between index/.style 2 args={
    x filter/.code={
        \ifnum\coordindex<#1\def\pgfmathresult{}\fi
        \ifnum\coordindex>#2\def\pgfmathresult{}\fi
    }
}}
\pgfplotstableread[col sep=comma,]{figures/data/experiment0_mode0.csv}\modezero
\pgfplotstableread[col sep=comma,]{figures/data/experiment0_mode2.csv}\modeone
\pgfplotstableread[col sep=comma,]{figures/data/experiment0_mode3.csv}\modetwo
\pgfplotstableread[col sep=comma,]{figures/data/experiment0_mode9.csv}\modethree
\begin{tikzpicture}
\tikzset{every node} = [font=\footnotesize]
\begin{axis} [
        name=mode0,
        ybar,
        bar width=.1,
        height=3cm,
        width=.59\linewidth,
        ymin=125,
        ymax=350,
        xticklabels={,,},
    ]
    \addplot[BlueBars] [
    ] table [
        x=addr, y=cycles, col sep=comma
    ] {\modezero};
    
    \node[draw,fill=white,anchor=north west] (m0) at (rel axis cs: 0,1) {(a)};
    \node[anchor=south east] (thresh0) at (axis cs: 123,200) {Threshold};
    \draw[-,dashed] (axis cs: -7,200) -- (thresh0.south east);
\end{axis}

\begin{axis} [
        name=mode1,
        ybar,
        bar width=.1,
        height=3cm,
        width=.59\linewidth,
        ymin=125,
        ymax=350,
        xticklabels={,,},
        yticklabels={,,},
        at={($(mode0.east) + (.5cm,0cm)$)},
        anchor=west,
    ]
    \addplot[BlueBars] [
    ] table [
        x=addr, y=cycles, col sep=comma
    ] {\modeone};
    
    \node[draw,fill=white,anchor=north west] (m1) at (rel axis cs: 0,1) {(b)};
    \node[anchor=south east] (thresh1) at (axis cs: 123,200) {Threshold};
    \draw[-,dashed] (axis cs: -7,200) -- (thresh0.south east);
\end{axis}

\begin{axis} [
        name=mode2,
        ybar,
        bar width=.1,
        height=3cm,
        width=.59\linewidth,
        ymin=125,
        ymax=350,
        at={($(mode0.south) + (0cm,-.5cm)$)},
        anchor=north,
    ]
    \addplot[BlueBars] [
        % select coords between index={0}{116}
    ] table [
        x=addr, y=cycles, col sep=comma
    ] {\modetwo};
    
    \node[draw,fill=white,anchor=north west] (m2) at (rel axis cs: 0,1) {(c)};
    \node[anchor=south east] (thresh2) at (axis cs: 123,200) {Threshold};
    \draw[-,dashed] (axis cs: -7,200) -- (thresh0.south east);
\end{axis}

\begin{axis} [
        name=mode3,
        ybar,
        bar width=.1,
        height=3cm,
        width=.59\linewidth,
        ymin=125,
        ymax=350,
        yticklabels={,,},
        at={($(mode1.south) + (0cm,-.5cm)$)},
        anchor=north,
    ]
    \addplot[BlueBars] [
    ] table [
        x=addr, y=cycles, col sep=comma
    ] {\modethree};
    
    \node[draw,fill=white,anchor=north west] (m3) at (rel axis cs: 0,1) {(d)};
    \node[anchor=south east] (thresh3) at (axis cs: 123,200) {Threshold};
    \draw[-,dashed] (axis cs: -7,200) -- (thresh0.south east);
\end{axis}

\path (mode2.south) -- node[midway,below,yshift=-.5cm]          {Eviction Set Address}    (mode3.south);
\path (mode0.west)  -- node[midway,above,rotate=90,yshift=.5cm] {Latency (\#FPGA Cycles)} (mode2.west);
\end{tikzpicture}
    \caption{Measurements for the conducted experiments with the SQL database. During measurement (a), the test app did not run any query. The queries run in measurements (b) - (d) returned no, one and 409600 rows of data from the database. It is clearly visible that the SQL queries leave a footprint in the IOTLB.}
    \label{fig:sql-exp0}
    \vspace*{-10pt}
\end{figure}
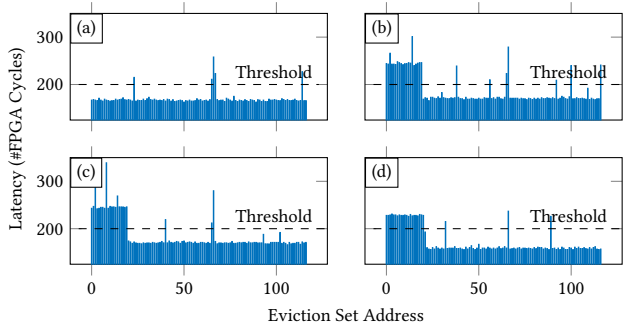

\autoref{fig:sql-exp0} (b) - (d) show probe measurements for queries returning no, one and 409,600 rows\footnote{One row in our case contains 36 bytes of data.} of data from the database.
During the measurement shown in \autoref{fig:sql-exp0} (a), no query was executed on the GPU.
The separate access times for each eviction set address are plotted along the x-axis.
The y-axis shows the measured latency for this address.
Clearly, the GPU leaves a footprint in the IOTLB when it computes an SQL query.
But, there is no measurable difference between the queries even if their results significantly differ in size.

Changing the test app to probe the eviction set while the SQL query executes on the GPU shows that the observable activity in the IOTLB is similar for all queries over time, besides the fact that queries with larger results produce longer traces as it takes longer to compute the result. Interestingly, the activity in the IOTLB happens towards the beginning of the query's computation.
At the time where the computed result is sent back to the CPU, there is no activity in the IOTLB.
This is easily explained by the way CUDA realizes the data transfer of the result from the GPU to the CPU: it uses MMIO\footnote{The CPU initializes the data transfer.} instead of DMA\footnote{The peripheral initializes the transfer.}.
We verified the explanation by inspecting the PCIe performance counters with the PCM tools\footnote{\texttt{pcm-pcie} -- \url{https://github.com/opcm/pcm}}.
The performance counters showed an increased amount of MMIO read requests that in total match the size of the returned result.

%%%%%%%%%%%%%%%%%%%%%%%%%%%%%%%%%%%%%%%%%%%%%%%%%%%%%%%%%%%%%%%%%%%%%%%%%%%%%%%%%%%%%%%
\revision{
\subsection{Side-Channel Impact}
\label{sec:sc-impact}
}
%%%%%%%%%%%%%%%%%%%%%%%%%%%%%%%%%%%%%%%%%%%%%%%%%%%%%%%%%%%%%%%%%%%%%%%%%%%%%%%%%%%%%%%
\revision{
So far, the observed leakage introduced by the IOTLB is mostly limited to a single bit describing whether a neighboring accelerator is in use or not. This is caused by two facts:
}

\revision{
\emph{(a)} Controlling an accelerator via MMIO rather than through DMA is a common usage model and limits the attack surface for IOTLB-based side-channel attacks because the CPU performs the address translation in the CPU's MMU instead of the GPU translating addresses via the IOMMU.
}

\revision{
\emph{(b)} Current PCIe devices usually perform DMA as bulk transfers, thereby limiting the overall PCIe protocol overhead. Loading data in a bulk transfer into device memory, computing on the data locally and eventually transferring the result back to the main memory in a bulk transfer means that no data-dependent access patterns -- which would leak information -- are observable in general.
}

\revision{
The two facts mentioned will likely change in the near future as PCIe 5.0 is rolled-out and Compute eXpress Link (CXL) is introduced.
\footnote{AMD CPUs and Intel FPGAs supporting CXL are already available. Intel plans rolling out compatible CPUs in the beginning of 2023~\cite{amd2022cxl,shilov2022cxl}.}
PCIe 5.0 reaches transfer speeds that are comparable with CPU main memory accesses.
This may lead device developers to include smaller memory on their devices and in turn access the main memory more often.
Furthermore, CXL features a coherency protocol that streamlines caching between main memory and PCIe device memory.
Again, this will lead device and driver developers to change from bulk transfers to more fine-grained data-dependent DMA accesses.
}

\revision{
In addition, FPGA vendors keep pushing for FPGA devices being the first-class compute device in a system while the CPU is merely used to manage the system and provide the FPGA with (increasingly sensitive) data.
Therefore, while the described side-channel is not yet very dangerous at the time of writing, it will become important in the near future. We highlight the side-channels existence and relevance \emph{before} widespread deployment of CXL and PCIe 5.
}

%%%%%%%%%%%%%%%%%%%%%%%%%%%%%%%%
\section{Covert Channels}
\label{sec:covert-channel}
%%%%%%%%%%%%%%%%%%%%%%%%%%%%%%%%
After identifying the IOTLB leakage and different ways to trigger and observe it, we now use our knowledge to construct two covert channels to prove the practicality of the channel \revision{with threat models \emph{2u} and \emph{2k}}.
The first channel is constructed between two peripherals and requires user privileges and DMA access to pages in main memory \revision{(model \emph{2u})}.
This channel could be implemented between two virtual machines, each with control of a DMA-enabled peripheral, such as Amazon's F1 FPGA instances or various GPU-enabled EC2 instances, as long as the two instances' peripherals share an IOMMU.
The performance of the covert channel can be improved if the receiver has root access on the host.
The second channel is unidirectional from CPU to peripheral and requires the sender to have root access to the host machine \revision{(model \emph{2k})}, thereby mostly serving as a proof of concept.
%We describe the second channel in \autoref{sec:cpu-fpga-covert-channel} in the appendix.
For both channels, the receiver must be able to measure time,~e.\,g.~through precise internal timers or high-speed network connection with external timers. This is the case for,~e.\,g.~GPUs~\cite{frigo2018gpu}, NICs and FPGAs.
\revision{All experiments in this section were run on the a10l system.}

\begin{table*}[t]
  \caption{Throughput and error rate for the covert channels tested on the \emph{a10l} system. For the peripheral-peripheral channel, sender and receiver are perfectly synchronous. The channel itself is very reliable which leads to nearly no errors. The throughput depends on the number of 1-bits in the message as each 1-bit is encoded into running a SQL-query on the sender peripheral which takes a rather long time of 0.3 seconds. For the CPU-peripheral channel, sender and receiver are not perfectly synchronous which leads to the rather high error rate. The throughput is limited by the speed of the CPU flushing the IOTLB. For both channels, plain bits were sent without encoding.}
  \label{tab:cc_stats}
  \centering
  \begin{tabular}{lllllcrrp{3cm}}
    \toprule
            & Sender     & Receiver   & Method    & Environment  & & Throughput & Error rate  & Content of message   \\
    \midrule                                                                            
    \multirow{4}{*}{\autoref{sec:peripheral-covert-channel}}
    & \multirow{4}{*}{Peripheral} & \multirow{4}{*}{Peripheral}
    & \multirow{4}{*}{Prime+Probe} & \multirow{4}{*}{Bare metal (cf. Fig.~\ref{fig:sql_test_stack})}
    & \begin{imageonly}\ldelim\{ {4}{*}\end{imageonly}                                           %[]
    &   3.4 bps  &     0\%     & All 1s \\
            &            &            &           &              & &   6.65 bps &     0\%     & Even mix of 1s and 0s \\
            &            &            &           &              & & 246.15 bps &   0.1\%     & All 0s \\
            &            &            &           &              & &   7.58 bps &     0\%     & ASCII-encoded text  \\
    \autoref{sec:cpu-fpga-covert-channel}
            & CPU        & Peripheral & Flush+Reload & Bare metal (cf. Fig.~\ref{fig:sql_test_stack}) & &  15023 bps & 30.09\%    & \textit{Performance not dependent on content} \\
    \bottomrule
  \end{tabular}
\end{table*}

%%%%%%%%%%%%%%%%%%%%%%%%%%%%%%%%%%%%%%%%%%%%%%%%%%
\subsection{Covert Channel between Peripherals}
\label{sec:peripheral-covert-channel}
%%%%%%%%%%%%%%%%%%%%%%%%%%%%%%%%%%%%%%%%%%%%%%%%%%
\begin{figure}[t]
  \centering
  \begin{subfigure}[t]{.9\linewidth}
    \centering
    \pgfplotsset{
    BlueBars/.style={
        draw=plotblue,
        fill=plotblue,
        bar width=0.25
    },
}
\pgfplotsset{select coords between index/.style 2 args={
    x filter/.code={
        \ifnum\coordindex<#1\def\pgfmathresult{}\fi
        \ifnum\coordindex>#2\def\pgfmathresult{}\fi
    }
}}

\pgfplotstableread[col sep=comma,]{figures/data/cc_gpu_fpga_Hello-World.csv}\datatable
\begin{tikzpicture}
\tikzset{every node} = [font=\footnotesize]
\begin{axis} [
        ybar,
        bar width=.1,
        height=3.5cm,
        width=\linewidth,
        ymin=0,
        ymax=29,
        xtick={0,8,16,24,32,40},
        xlabel={Bit \#},
        ylabel={\# IOTLB Misses},
    ]
    \addplot[BlueBars] [
        select coords between index={0}{39}
    ] table [
        x=bit, y=misses, col sep=comma
    ] {\datatable};
    
    \node[] at (axis cs:    0,23) {0};
    \node[] at (axis cs:    1,23) {1};
    \node[] at (axis cs:    2,23) {0};
    \node[] at (axis cs:    3,23) {0};
    \node[] at (axis cs:    4,23) {1};
    \node[] at (axis cs:    5,23) {0};
    \node[] at (axis cs:    6,23) {0};
    \node[] at (axis cs:    7,23) {0};
    \node[] at (axis cs:  3.5,27) {H};
    \draw [decorate,decoration={calligraphic brace}] (axis cs: -.25,24.5) -- (7.25,24.5);
    
    \node[] at (axis cs:    8,23) {0};
    \node[] at (axis cs:    9,23) {1};
    \node[] at (axis cs:   10,23) {1};
    \node[] at (axis cs:   11,23) {0};
    \node[] at (axis cs:   12,23) {0};
    \node[] at (axis cs:   13,23) {1};
    \node[] at (axis cs:   14,23) {0};
    \node[] at (axis cs:   15,23) {1};
    \node[] at (axis cs: 11.5,27) {e};
    \draw [decorate,decoration={calligraphic brace}] (axis cs: 7.75,24.5) -- (15.25,24.5);
    
    \node[] at (axis cs:   16,23) {0};
    \node[] at (axis cs:   17,23) {1};
    \node[] at (axis cs:   18,23) {1};
    \node[] at (axis cs:   19,23) {0};
    \node[] at (axis cs:   20,23) {1};
    \node[] at (axis cs:   21,23) {1};
    \node[] at (axis cs:   22,23) {0};
    \node[] at (axis cs:   23,23) {0};
    \node[] at (axis cs: 19.5,27) {l};
    \draw [decorate,decoration={calligraphic brace}] (axis cs: 15.75,24.5) -- (23.25,24.5);
    
    \node[] at (axis cs:   24,23) {0};
    \node[] at (axis cs:   25,23) {1};
    \node[] at (axis cs:   26,23) {1};
    \node[] at (axis cs:   27,23) {0};
    \node[] at (axis cs:   28,23) {1};
    \node[] at (axis cs:   29,23) {1};
    \node[] at (axis cs:   30,23) {0};
    \node[] at (axis cs:   31,23) {0};
    \node[] at (axis cs: 27.5,27) {l};
    \draw [decorate,decoration={calligraphic brace}] (axis cs: 23.75,24.5) -- (31.25,24.5);
    
    \node[] at (axis cs:   32,23) {0};
    \node[] at (axis cs:   33,23) {1};
    \node[] at (axis cs:   34,23) {1};
    \node[] at (axis cs:   35,23) {0};
    \node[] at (axis cs:   36,23) {1};
    \node[] at (axis cs:   37,23) {1};
    \node[] at (axis cs:   38,23) {1};
    \node[] at (axis cs:   39,23) {1};
    \node[] at (axis cs: 35.5,27) {o};
    \draw [decorate,decoration={calligraphic brace}] (axis cs: 31.75,24.5) -- (39.25,24.5);
\end{axis}
\end{tikzpicture}
    \caption{Scenario as in \autoref{fig:sql_test_stack}; big endian transmission. The FPGA uses an eviction set that was constructed using IOTLB flushes. This results in very reliable eviction sets and in turn a reliable transmission.}
    \label{fig:cc-gpu-fpga}
  \end{subfigure}\\
  \begin{subfigure}[t]{.9\linewidth}
    \centering
    \pgfplotsset{
    BlueBars/.style={
        draw=plotblue,
        fill=plotblue,
        bar width=0.25
    },
}
\pgfplotsset{select coords between index/.style 2 args={
    x filter/.code={
        \ifnum\coordindex<#1\def\pgfmathresult{}\fi
        \ifnum\coordindex>#2\def\pgfmathresult{}\fi
    }
}}

\pgfplotstableread[col sep=comma,]{figures/data/cc_vm_gpu_fpga_Hello.csv}\datatable
\begin{tikzpicture}
\tikzset{every node} = [font=\footnotesize]
\begin{axis} [
        ybar,
        bar width=.1,
        height=3.5cm,
        width=\linewidth,
        ymin=0,
        ymax=37,
        xtick={0,8,16,24,32,40},
        xlabel={Bit \#},
        ylabel={\# IOTLB Misses},
    ]
    \addplot[BlueBars] [
        select coords between index={0}{39}
    ] table [
        x=bit, y=misses, col sep=comma
    ] {\datatable};
    
    \node[] at (axis cs:    0,29) {0};
    \node[] at (axis cs:    1,29) {0};
    \node[] at (axis cs:    2,29) {0};
    \node[] at (axis cs:    3,29) {1};
    \node[] at (axis cs:    4,29) {0};
    \node[] at (axis cs:    5,29) {0};
    \node[] at (axis cs:    6,29) {1};
    \node[] at (axis cs:    7,29) {0};
    \node[] at (axis cs:  3.5,34) {H};
    \draw [decorate,decoration={calligraphic brace}] (axis cs: -.25,31) -- (7.25,31);
    
    \node[] at (axis cs:    8,29) {1};
    \node[] at (axis cs:    9,29) {0};
    \node[] at (axis cs:   10,29) {1};
    \node[] at (axis cs:   11,29) {0};
    \node[] at (axis cs:   12,29) {0};
    \node[] at (axis cs:   13,29) {1};
    \node[] at (axis cs:   14,29) {1};
    \node[] at (axis cs:   15,29) {0};
    \node[] at (axis cs: 11.5,34) {e};
    \draw [decorate,decoration={calligraphic brace}] (axis cs: 7.75,31) -- (15.25,31);
    
    \node[] at (axis cs:   16,29) {0};
    \node[] at (axis cs:   17,29) {0};
    \node[] at (axis cs:   18,29) {1};
    \node[] at (axis cs:   19,29) {1};
    \node[] at (axis cs:   20,29) {0};
    \node[] at (axis cs:   21,29) {1};
    \node[] at (axis cs:   22,29) {1};
    \node[] at (axis cs:   23,29) {0};
    \node[] at (axis cs: 19.5,34) {l};
    \draw [decorate,decoration={calligraphic brace}] (axis cs: 15.75,31) -- (23.25,31);
    
    \node[] at (axis cs:   24,29) {0};
    \node[] at (axis cs:   25,29) {0};
    \node[] at (axis cs:   26,29) {1};
    \node[] at (axis cs:   27,29) {1};
    \node[] at (axis cs:   28,29) {0};
    \node[] at (axis cs:   29,29) {1};
    \node[] at (axis cs:   30,29) {1};
    \node[] at (axis cs:   31,29) {0};
    \node[] at (axis cs: 27.5,34) {l};
    \draw [decorate,decoration={calligraphic brace}] (axis cs: 23.75,31) -- (31.25,31);
    
    \node[] at (axis cs:   32,29) {1};
    \node[] at (axis cs:   33,29) {1};
    \node[] at (axis cs:   34,29) {1};
    \node[] at (axis cs:   35,29) {1};
    \node[] at (axis cs:   36,29) {0};
    \node[] at (axis cs:   37,29) {1};
    \node[] at (axis cs:   38,29) {1};
    \node[] at (axis cs:   39,29) {0};
    \node[] at (axis cs: 35.5,34) {o};
    \draw [decorate,decoration={calligraphic brace}] (axis cs: 31.75,31) -- (39.25,31);
    
    \draw [dashed] (axis cs: -5,16.75) -- (axis cs: 45,16.75);
\end{axis}
\end{tikzpicture}
    \caption{Scenario as in \autoref{fig:sql_vm_test_stack}; little endian transmission. The FPGA uses eviction sets constructed \emph{without} IOTLB flushes. TEven though the transmission is free of errors, buit turns out to be more noisy.}
    \label{fig:cc-vm-gpu-fpga}
  \end{subfigure}
  \caption{Peripheral to peripheral covert channel transmissions of t. The message ``Hello'' was sent in big endian format.}
  \label{fig:cc-transmission}
\end{figure}
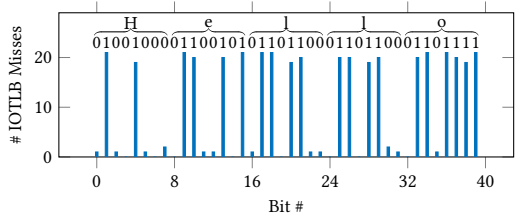
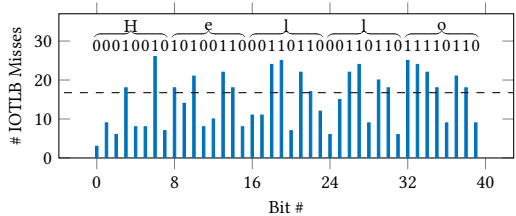

Another research question is whether two peripherals can use the IOTLB to construct a covert channel between each other.
To answer this question, we co-locate the Arria 10 with the Tesla T4.
Our goal is to use the footprint that an SQL query computed on the GPU leaves in the IOTLB to send information to the FPGA.
Such a covert channel exists in a scenario where the sender uses a website that, depending on the actions performed on the website, runs SQL queries on a GPU-accelerated database.
The sender can then exploit the website to send information to the co-located FPGA.

We prepare a10l as shown in \autoref{fig:sql_test_stack}.
The sender encodes a one into running an SQL query and running no query encodes a zero.
The receiver uses the \texttt{iotlb\_pnp} hardware function on the FPGA to monitor the IOTLB using the Prime+Probe technique.
Each SQL query evicts 18-20 entries of the receiver's eviction set (cf. \autoref{fig:sql-exp0} (b) - (d)).
A plot of the number of IOTLB misses measured during message transmission is given in \autoref{fig:cc-gpu-fpga}.
We found that basically no errors occur if sender and receiver are synchronized.
This means that the channel is nearly free of bit-flip errors.
If perfect synchronization is not achievable, the channel suffers from insertion and deletion errors.
In this case techniques from~\cite{maurice2017hello} can be applied to overcome these errors.
The channel's throughput highly depends on the number of one bits in the message.
This is because the execution time of a single SQL query takes about 0.3 seconds.
\autoref{tab:cc_stats} shows more detailed measurements for different 0-1-ratios in the message that is transferred over the covert channel.
Of course, a GPU application optimized for acting as a sender in this scenario would allow us to increase the bandwidth of the channel.

\begin{figure}[t]
    \centering
    \begin{tikzpicture}[
    node distance=1.2cm,
    alignc/.style={align=center},
    alignl/.style={align=left},
    alignr/.style={align=right},
    small dim/.style={minimum height=.5cm, minimum width=3cm, text width=2.75cm},
    wide dim/.style={minimum height=.5cm, minimum width=6cm, text width=5.75cm},
    medium wide dim/.style={minimum height=.5cm, minimum width=9cm, text width=8.75cm},
    very wide dim/.style={minimum height=.5cm, minimum width=12cm, text width=11.75cm},
    green/.style={draw=nicegreenborder, fill=nicegreenfill, very thick},
    blue/.style={draw=niceblueborder, fill=nicebluefill, very thick},
    red/.style={draw=niceredborder, fill=niceredfill, text=white, very thick},
    orange/.style={draw=niceorangeborder, fill=niceorangefill, very thick},
    yellow/.style={draw=niceyellowborder, fill=niceyellowfill, very thick},
]

\tikzset{every node} = [font=\footnotesize]

\node[wide dim, alignc, green] (cpu) {CPU};
\node[wide dim, alignc, green, anchor=north]      (iommu) at (cpu.south) {IOMMU};

\node[small dim, alignc, red, anchor=south west]  (vfio0) at (cpu.north) {vfio-pci};
\node[small dim, alignc, blue, anchor=south east] (vfio1) at (cpu.north) {vfio-pci};

\node[small dim, alignc, yellow, anchor=south]  (kvm0) at (vfio0.north) {KVM};
\node[small dim, alignc, yellow, anchor=south] (kvm1) at (vfio1.north) {KVM};

\node[small dim, alignc, red, anchor=south]  (fpga-driver) at (kvm0.north) {intel-fpga-pci};
\node[small dim, alignc, blue, anchor=south] (gpu-driver) at (kvm1.north) {cuda};

\node[small dim, alignc, orange, anchor=south] (sqldb) at (gpu-driver.north) {Omnisci DB};
\node[small dim, alignc, orange, anchor=south] (attacker) at (fpga-driver.north) {Receiver};

\node[small dim, alignc, blue, below=1cm of iommu.south, anchor=east] (gpu)  {NVIDIA T4 GPU};
\node[small dim, alignc, red, below=1cm of iommu.south, anchor=west]  (fpga) {Arria 10 FPGA};

\node[text width=2cm, alignr, anchor=east] at (sqldb.west) {applications};
\node[text width=2cm, alignr, anchor=east] at (gpu-driver.west) {drivers in VM};
\node[text width=2cm, alignr, anchor=east] at (kvm1.west) {hypervisor};
\node[text width=2cm, alignr, anchor=east] at (vfio1.west) {drivers};
\node[text width=2cm, alignr, anchor=east] at (cpu.south west) {internal hardware};
\path (iommu.west) -- node [midway, text width=2cm, anchor=east, alignr] {PCI-e linkage} (gpu.west);
\node[text width=2cm, alignr, anchor=east] at (gpu.west) {peripheral devices};

\path[<->, draw, dashed, very thick] (gpu.north) -- (gpu.north |- iommu.south);
\path[<->, draw, dashed, very thick] (fpga.north) -- (fpga.north |- iommu.south);
\path[<->, draw, dotted, very thick] (vfio1.south) -- (gpu.north |- iommu.south);
\path[<->, draw, dotted, very thick] (vfio0.south) -- (fpga.north |- iommu.south);
\path[draw, red, dotted, very thick] ([yshift=0pt]gpu.north |- iommu.center) edge[->, bend left=15] ([yshift=0pt]fpga.north |- iommu.center);
\path[<->, draw, very thick] ([yshift=-5pt]sqldb.center) -- ([yshift=-5pt]attacker.center);
\end{tikzpicture}
    \caption{Stack diagram of the GPU accelerated SQL database covert channel across virtual machines.}
    \label{fig:sql_vm_test_stack}
\end{figure}
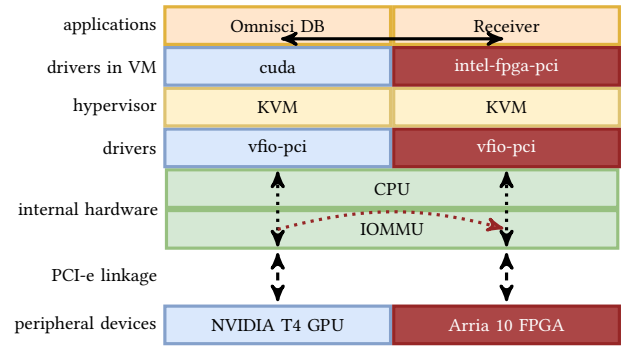

For the previous test, the eviction set used by the FPGA was constructed with IOTLB flushes to work with eviction sets of optimal reliability.
As mentioned earlier, IOTLB flushes require kernel privileges on commodity host Linux systems.
User-level receivers or receivers located in virtual machines have to use the less reliable eviction sets constructed without IOTLB flushes.
As can be seen in \autoref{fig:cc-vm-gpu-fpga}, this results in more noise in the measurements.
The depicted transmission is still free of errors but some bits are at the edge of being falsely classified.
To overcome potential bitflip errors, error detection mechanisms like CRC codes or error correction codes like Hadamard codes can be applied~\cite{maurice2017hello}.
The presented covert channel works between any two peripherals that use DMA to access the main memory.
For the receiver, the accessible memory region needs to be sufficiently large to allow for eviction set construction.
Additionally, the receiver needs a mechanism to measure the memory access latency.
Programmable or configurable peripherals like FPGAs or GPUs will meet both receiver requirements even in the most stringent cloud environments if bare metal instances are available for rent.
An FPGA or GPU sender has fine-grained control of the channel, but a more opaque sender like a smart NIC or PCIe-enabled storage device could work as a sender, albeit more likely to be noisy or unreliable.

Peripherals that manage secrets and perform DMAs depending on the value of the secret must be aware that neighboring devices connected to the same IOMMU may be able to observe their access patterns.
This is especially true for peripherals where the programming model assumes unified memory that abstracts separate physical memory locations like device and system memory away from the developer as in this case the leaking DMA may occur without the knowledge of the developer.
As of today, data-dependent DMA is used seldomly due to the overhead that renders it inefficient.
\revision{
But we expect this behavior to change with the introduction of PCIe 5.0 and CXL as mentioned in earlier sections.
}
%%%%%%%%%%%%%%%%%%%%%%%%%%%%%%%%%%%%%%%%%%%%%%%%%%%%
\revision{
\subsection{Covert Channel from CPU to Peripheral}
\label{sec:cpu-fpga-covert-channel}
}
%%%%%%%%%%%%%%%%%%%%%%%%%%%%%%%%%%%%%%%%%%%%%%%%%%%%
The CPU is very limited in interacting with the IOTLB directly.
Because the IOMMU translates addresses for peripherals only, memory accesses from the CPU do not interfere with the IOTLB.
The only way for the CPU to interfere with the IOTLB is by changing page table entries or instructing the IOMMU to flush certain (or all) entries in the IOTLB. Usually, only the OS, hypervisor or VMM issues page table changes or IOTLB flushes, which is why the Linux kernel does not provide an interface for flushing the IOTLB to userland.
To overcome this problem, we load a self-developed kernel module that exposes a IOTLB flush API to our test application.
An overview of our system setup for this covert channel is given in \autoref{fig:sql_test_stack}.

Since a peripheral can distinguish IOTLB hits from misses, flushing the IOTLB allows the CPU to send information covertly to peripherals.
A global IOTLB flush takes \unit[17]{$\mu$s} on average.
Flushing all entries from the IOTLB encodes a \texttt{1} and sleeping for \unit[17]{$\mu$s} encodes a \texttt{0}.
As the receiver we use the \texttt{iotlb\_pnp} hardware module described in \autoref{sec:hardware-design}.
The hardware function is programmed to continuously probe a fixed target address.
Whenever a probe reports a slow access, a \texttt{1} is received.
Otherwise, the hardware receives a \texttt{0}.
We implement the covert channel in a trivial way without applying any encoding for error correction or synchronization.
Because a memory access from the FPGA running at \unit[200]{MHz} only takes around \unit[1]{$\mu$s} we roughly synchronize the FPGA with the CPU by making the FPGA \texttt{wait} for a certain amount of cycles.
We determined the number of cycles to wait by repeatedly flushing the IOTLB and increasing the number of wait cycles until all FPGA memory accesses are slow.
After this very rough synchronization step, a message of $2^{16}-1$ bits generated by a linear feedback-shift register is transmitted to measure throughput and error rates.
The result is given in \autoref{tab:cc_stats}.
As can be seen, this basic covert channel without further optimizations already achieves a throughput of around \unit[15]{kBit/s}.
The error rate is 30\% which can be improved significantly by applying error-correction and error-handling techniques as e.\,g. described in \cite{maurice2017hello}.

Because so far the covert channel only offers communication in one direction, we tried to improve the channel to offer bi-directional message transfer.
To do so we checked the timing behavior of flushing the IOTLB.
The \texttt{clflush} instruction on x86 CPUs has a data-dependent execution time~\cite{gruss2016flush}.
In our case, a data-dependency of the flush time on IOTLB entries would allow us to construct the reverse covert channel.
However, our experiments show no measurable timing behavior of the flush that can be related to the usage of the IOTLB; an IOTLB flush takes around \unit[17]{$\mu$s} independent of FPGA memory accesses before or even during the flush. The latency is also independent from whether only addresses of a certain peripheral or all entries of the IOTLB are flushed. However, peripheral-to-CPU covert channels based on the CPU cache do exist \cite{weissman2020jackhammer}.

The demonstrated covert channel is reliable without applying special synchronization, error-correction, or error-detection techniques.
However, only peripherals can act as the receiver while the CPU is limited to the role of the sender.
Also, with the standard IOMMU drivers in Linux, the sending process is required to run kernel-level code to perform IOTLB flushes.
A privileged device driver that flushes the IOTLB under certain circumstances may expose this flushing capability to an unprivileged user.
Device drivers that make extensive use of IOTLB flushes may also be vulnerable to a side-channel attack from an untrusted peripheral device that monitors the IOTLB for flushes.
For example, a driver developer may chose to include IOTLB flushes to remove traces of a trusted peripheral's activity for security; however, the timing between flushes could leak information about the operation of an application using that peripheral.

%%%%%%%%%%%%%%%%%%%%%%%%%%%%%%%%
\section{Countermeasures}
\label{sec:countermeasures}
%%%%%%%%%%%%%%%%%%%%%%%%%%%%%%%%
Like many microarchitectural attacks, there are a variety of defenses against IOTLB side-channels that can be implemented at nearly any level of a system. 
\revision{
We first present immediately available actions that can be taken by system administrators and cloud application developers, and then discuss defenses that can be built into future IOMMU architectures.
}

\subsection{\revision{Securing Existing Systems}}
In cases where multiple users who do not trust each other may use the same machine, ensuring that no two users (or no one user and the hypervisor) have access to peripherals on the same IOMMU hardware is sufficient to protect against IOTLB side-channel attacks. 
\revision{
On a Linux host, \texttt{/sys/class/iommu/} provides information on a system's IOMMU devices and the PCIe devices that use them~\cite{williamson2014iommu}.
Typically, systems have several IOMMU devices, each of which is linked to a few PCIe endpoints, which may be internal PCIe devices or external devices plugged into PCIe slots on the motherboard. 
Endpoints cannot be reassigned to new IOMMUs, so ensuring full isolation may limit scaling capacity.%, particularly in cloud environments. 
For example, a CSP could not use a motherboard with eight full-size, full-speed PCIe slots managed in pairs by four IOMMUs to provide eight fully isolated single-GPU cloud instances, even though eight GPUs fit in the PCIe slots of the system.
}

\revision{
On the \emph{application level}, code and hardware involved in data dependent computation can rely on constant time algorithms with constant memory access patterns, so
}
no information about the operations is leaked through the IOTLB. For cryptographic implementations this is a common technique but for database systems constant memory access patterns and timings are not easily achieved. Private Information Retrieval (PIR) protocols~\cite{chor1998pir,kushilevitz1997replicationnotneeded} can be a solution, but modern implementations\footnote{e.\,g. \url{https://github.com/ReverseControl/MuchPIR}} usually only support index queries. Recent attempts~\cite{hayata2020pirrangequeries} to also support range queries may still leak information about the response size.

\revision{
\emph{A hypervisor} can enable Address Translation Services (ATS)~\cite{pcisig2009ats} for a peripheral to remove all of its traces from its IOTLB.
}
Address Translation Services (ATS) allows a device to maintain and use a local on-device TLB for address translation and selectively bypass IOMMU translation. Since locally-translated requests are not translated by the IOMMU, they do not leave any trace in the IOTLB.
\revision{However, devices must specifically support ATS to use it, and furthermore, allowing ATS for untrusted devices is not advisable.
}
ATS allows a device to provide \emph{any} physical address as part of a DMA request and mark it as ``translated''.
Malicious devices may exploit ATS for unrestricted physical memory access~\cite{markettos2019thunderclap}.
Therefore, ATS must only be allowed for trusted devices.

\revision{
Hypervisors can also achieve a separation of the IOTLB between mutually untrusted tenants by IOTLB partitioning.
}
For set-associative IOTLBs, set partitioning can be done by the hypervisor in software by only allocating I/O virtual addresses of sets to each tenant~\cite{ye2014coloris}. However, set-based partitioning may not work with peripherals that rely on the address space being contiguous.

\subsection{\revision{Securing Future IOMMUs}}
\revision{
If \emph{hardware modifications} are a viable option to implement countermeasures, then way-based partitioning is another option.
}
It needs to be supported by the IOMMU hardware so that the hypervisor can map each address of a thread to a fixed number of ways like is possible with Intel CAT~\cite{intel2016cat,liu2016catalyst} for CPU-internal caches.

Future IOMMUs could include support for flagging a page translation as uncacheable. This would ensure that it is never stored in the IOTLB and that the use of that page would never affect the IOTLB state, so it would be invisible to any side-channel attack. However, all accesses to that page would be as slow as IOTLB misses, increasing latency and likely reducing maximum throughput.

%%%%%%%%%%%%%%%%%%%%%%%%%%%%%%%%
\section{Conclusion}
\label{sec:conclusion}
%%%%%%%%%%%%%%%%%%%%%%%%%%%%%%%%
State-of-the-art cloud environments use direct memory access managed by IOMMUs to offer high speed, low latency, and isolated memory access to an increasingly wide variety of peripherals.
These peripherals support and accelerate many types of applications and virtual hardware functions, including those that perform secure operations or handle sensitive data.
In this paper we demonstrated a new side-channel attack against IOTLBs in such IOMMUs that works across virtual environments and threatens cloud tenants.
We developed a new eviction set finding algorithm that works without prior assumptions of cache or TLB organization and a hardware module for an FPGA that implements the fundamentals necessary to exploit the IOTLB side-channel.
We used these tools to record a side-channel trace from a GPU running a database acceleration library.
The results prove that the IOTLB can be used as a side-channel to spy on co-located devices.
We highlight this fact by showing a very reliable covert channel from the GPU to the FPGA where we use the database application running on the GPU to encode messages into the GPU's system memory access patterns.
\revision{
While we acknowledge the limitations of the IOTLB channel with current hardware and applications, we argue that with the upcoming PCIe 5.0 and CXL standards, IOMMU usage patterns will change and fine-grained IOTLB side-channel attacks will become practical.
To overcome the threat of the side-channel, we suggest a variety of countermeasures that can be implemented on different system levels ranging from hardware modifications up to the implementation of applications.
}
Many of these countermeasures fully eliminate the threat of IOTLB side-channels, but at the same time reduce the speed of peripherals or scalability of the systems that host them.
Therefore, when designing or choosing hardware for large-scale, high-performance, secure services, IOTLB threats must be acknowledged and IOTLB isolation measures must be carefully considered for the specific needs of the system. 
Furthermore, when designing security-critical peripherals or security-critical software or firmware that makes use of peripherals, timing leakages from peripheral memory accesses must be addressed with constant-time design practices.

%\ifdefined\PUBLICATION
%%%%%%%%%%%%%%%%%%%%%%%%%%%%%%%%
\begin{acks}
%%%%%%%%%%%%%%%%%%%%%%%%%%%%%%%%
Worcester Polytechnic Institute is located on the traditional land of the Nipmuc people.
This research received partial funding, hardware donations, and extremely useful advice from Intel and its employees. 
We especially thank Alpa Trivedi, Sayak Ray, and Thomas Unterluggauer from Intel as well as our reviewers for their advice and comments.
This research was also partially funded by: the German Research Foundation (DFG) grant 456967092 (SecFShare); the German Federal Ministry of Education and Research (BMBF) grant VE-Jupiter (FKZ 16ME0234); the National Science Foundation (NSF) grants CNS-1814406, and CNS-2026913.
\end{acks}
%\fi

%%%%%%%%%%%%%%%%%%%%%%%%%%%%%%%%
% References
%%%%%%%%%%%%%%%%%%%%%%%%%%%%%%%%
\bibliographystyle{ACM-Reference-Format}
\bibliography{reference}

%%%%%%%%%%%%%%%%%%%%%%%%%%%%%%%%
\appendix
%%%%%%%%%%%%%%%%%%%%%%%%%%%%%%%%
%\revision{
\section{Appendix}
%}

%%%%%%%%%%%%%%%%%%%%%%%%%%%%%%%%%%%%%%%%%%%%%%%%%%%%
\revision{
\subsection{Initial IOTLB Organization Hypothesis}
\label{sec:initial_hypothesis}
}
%%%%%%%%%%%%%%%%%%%%%%%%%%%%%%%%%%%%%%%%%%%%%%%%%%%%
Initially, we hypothesized that the IOTLB would be organized like the CPU TLBs reverse-engineered in \cite{gras2018translation}, with $2^s$ sets where $s$ is an integer, some small number of ways per set, and a set mapping algorithm wherein the lowest $s$ bits of the page address select the set number or some other combination of various bits of the page address forms the set number that the page is associated with.
Initial experiments on all three systems showed that 128-address eviction sets of any randomly allocated pages reliably evicted any other single page, so we hypothesized that the IOTLB was organized with 128 sets and 1 way.
We tested this hypothesized eviction set architecture in a scenario on \emph{a10l} where the FPGA used Prime+Probe to monitor an IOTLB that it shared with a network card.

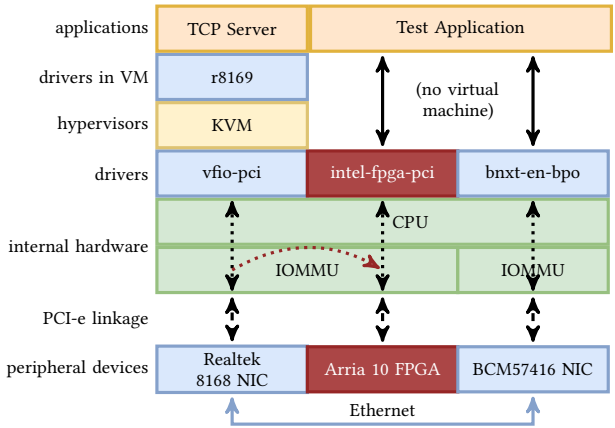
\begin{figure}[t]
    \centering
    \begin{tikzpicture}[
    node distance=1.2cm,
    alignc/.style={align=center},
    alignl/.style={align=left},
    alignr/.style={align=right},
    small dim/.style={minimum height=.6cm, minimum width=2cm, text width=1.75cm},
    wide dim/.style={minimum height=.6cm, minimum width=4cm, text width=3.75cm},
    medium wide dim/.style={minimum height=.6cm, minimum width=6cm, text width=5.75cm},
    very wide dim/.style={minimum height=.6cm, minimum width=8cm, text width=7.75cm},
    green/.style={draw=nicegreenborder, fill=nicegreenfill, very thick},
    blue/.style={draw=niceblueborder, fill=nicebluefill, very thick},
    red/.style={draw=niceredborder, fill=niceredfill, text=white, very thick},
    orange/.style={draw=niceorangeborder, fill=niceorangefill, very thick},
    yellow/.style={draw=niceyellowborder, fill=niceyellowfill, very thick},
]

\tikzset{every node} = [font=\footnotesize]

\node[medium wide dim, alignc, green] (cpu) {\quad\quad\quad CPU};
\node[wide dim, alignc, green, anchor=north west]      (iommu1) at (cpu.south west) {IOMMU};
\node[small dim, alignc, green, anchor=north east]      (iommu2) at (cpu.south east) {IOMMU};

\node[small dim, alignc, blue, anchor=south west] (nic-driver1) at (cpu.north west) {vfio-pci};
\node[small dim, alignc, red, anchor=south]  (fpga-driver) at (cpu.north) {intel-fpga-pci};
\node[small dim, alignc, blue, anchor=south east] (nic-driver2) at (cpu.north east) {bnxt-en-bpo};

\node[small dim, alignc, yellow, anchor=south] (kvm) at (nic-driver1.north) {KVM};
\node[small dim, alignc, blue, anchor=south] (vm-driver) at (kvm.north) {r8169};

\node[small dim, alignc, orange, anchor=south] (tcps) at (vm-driver.north) {TCP Server};
\node[wide dim, alignc, orange, anchor=west] (attacker) at (tcps.east) {Test Application};

\node[small dim, alignc, blue, below=1cm of iommu1.south west, anchor=west] (nic1) {Realtek 8168 NIC};
\node[small dim, alignc, red, below=1cm of iommu1.south east, anchor=east]  (fpga) {Arria 10 FPGA};
\node[small dim, alignc, blue, below=1cm of iommu2.south east, anchor=east] (nic2) {BCM57416 NIC};

\node[text width=2cm, alignr, anchor=east] at (tcps.west) {applications};
\node[text width=2cm, alignr, anchor=east] at (vm-driver.west) {drivers in VM};
\node[text width=2cm, alignr, anchor=east] at (kvm.west) {hypervisors};
\node[text width=2cm, alignr, anchor=east] at (nic-driver1.west) {drivers};
\node[text width=2cm, alignr, anchor=east] at (cpu.south west) {internal hardware};
\path (iommu1.west) -- node [midway, anchor=east, text width=2cm, alignr] {PCI-e linkage} (nic1.west);
\node[text width=2cm, alignr, anchor=east] at (nic1.west) {peripheral devices};

\path[<->, draw, dashed, very thick] (nic1.north) -- (nic1.north |- iommu1.south);
\path[<->, draw, dashed, very thick] (fpga.north) -- (fpga.north |- iommu1.south);
\path[<->, draw, dashed, very thick] (nic2.north) -- (nic2.north |- iommu2.south);

\path[<->, draw, dotted, very thick] (nic-driver1.south) -- (nic1.north |- iommu1.south);
\path[<->, draw, dotted, very thick] (fpga-driver.south) -- (fpga.north |- iommu1.south);
\path[<->, draw, dotted, very thick] (nic-driver2.south) -- (nic-driver2.north |- iommu2.south);

\path[<->, draw, very thick] (fpga-driver.north) -- node [midway] (tmp1) {} (fpga-driver.north |- attacker.south);
\path[<->, draw, very thick] (nic-driver2.north) -- node [midway] (tmp2) {} (nic-driver2.north |- attacker.south);
\path (tmp1) -- node [midway,text width=2cm,align=center] {(no virtual machine)} (tmp2);

\path[draw, red, dotted, very thick] ([yshift=0pt]nic1.north |- iommu1.center) edge[->, bend left=30] ([yshift=0pt]fpga.north |- iommu1.center);

\path[<->, draw=niceblueborder, very thick] (nic1.south) -- +(0,-.4) -| node [pos=.25, above] {Ethernet} (nic2.south);
\end{tikzpicture}
    \caption{Stack diagram of the network card side-channel test. The Realtek network interface card (NIC) is "passed through" to a virtual machine with the VFIO driver. The test application exchanges packets with the TCP server in the virtual machine over the ethernet connection between the two network cards; meanwhile, the FPGA (connected to the same IOMMU as the VM's network card) probes the IOTLB for traces of network activity.}
    \label{fig:tcp_test_stack}
\end{figure}

\autoref{fig:tcp_test_stack} shows the hardware and software setup for this test\revision{, an example of threat model \emph{1u}}. 
A virtual machine is configured with the IOMMU in a pass-through mode (Virtual Function I/O or VFIO) to allow a Realtek 8168 NIC direct access to the virtual environment, where it uses the standard r8169 drivers.
The test application runs directly on the host, and uses the Broadcom BCM57416 NIC to exchange packets with the Realtek NIC over ethernet.
The test application also manages our Prime+Probe hardware on the Arria 10 GX FPGA and uses it to collect IOTLB side-channel traces while the network is active.
The eviction sets used in the Prime+Probe tests are constructed under the assumption that the IOTLB contains 128 sets of one way each. 

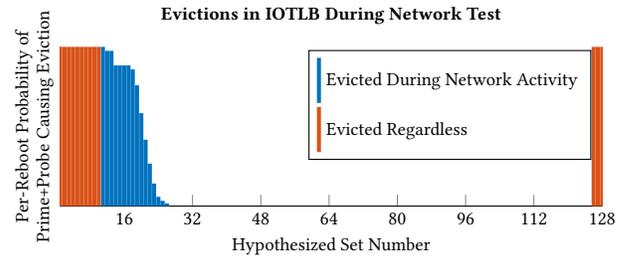
\begin{figure}[t]
    \centering
    % This file was created by matlab2tikz.
%
%The latest updates can be retrieved from
%  http://www.mathworks.com/matlabcentral/fileexchange/22022-matlab2tikz-matlab2tikz
%where you can also make suggestions and rate matlab2tikz.
%
\newcommand\axwidth{.8\linewidth}
\newcommand\barwidth{.004\linewidth} % 1/128 of axwidth
\pgfplotsset{compat=1.16,
    /pgfplots/vline legend/.style={
        legend image code/.code={
            \draw [mark repeat=2,mark phase=2,##1]
             plot coordinates {
            (0.3cm,-0.3cm) (0.3cm,0cm)    (0.3cm,0.3cm)
            };
}, },
}
\begin{tikzpicture}

\tikzset{every node} = [font=\footnotesize]

\begin{axis}[%
    width=.85\linewidth,
    height=.25*\linewidth,
    scale only axis,
    bar shift auto,
    xmin=1,
    xmax=128,
    xtick={ 16,  32,  48,  64,  80,  96, 112, 128},
    xminorticks=true,
    xlabel={Hypothesized Set Number},
    ymin=0,
    ymax=1,
    ytick=\empty,
    ylabel style={
        % font=\color{white!15!black},
        align=center
    },
    ylabel={Per-Reboot Probability of\\Prime+Probe Causing Eviction},
    axis background/.style={fill=white},
    title style={font=\bfseries},
    title={Evictions in IOTLB During Network Test},
    axis x line*=bottom,
    axis y line*=left,
    legend style={
        legend cell align=left,
        align=right,
        % draw=white!15!black
    }
]
\addplot[ycomb, draw=plotblue, line width=.00630\linewidth, vline legend] table[row sep=crcr] {%
11	1\\
12	0.975\\
13	0.975\\
14	0.883333333333333\\
15	0.883333333333333\\
16	0.883333333333333\\
17	0.883333333333333\\
18	0.858333333333333\\
19	0.758333333333333\\
20	0.583333333333333\\
21	0.416666666666667\\
22	0.266666666666667\\
23	0.141666666666667\\
24	0.0583333333333333\\
25	0.0333333333333333\\
26	0.0166666666666667\\
};
%\addplot[forget plot, color=white!15!black] table[row sep=crcr] {%
%1	0\\
%128	0\\
%};
\addlegendentry{Evicted During Network Activity}

\addplot[ycomb, draw=plotorange, line width=.00630\linewidth, vline legend] table[row sep=crcr] {%
1	1\\
2	1\\
3	1\\
4	1\\
5	1\\
6	1\\
7	1\\
8	1\\
9	1\\
10	1\\
126	1\\
127	1\\
128	1\\
};
%\addplot[forget plot, color=white!15!black] table[row sep=crcr] {%
%1	0\\
%128	0\\
%};
\addlegendentry{Evicted Regardless}

\end{axis}

%begin{axis}[%
%idth=2.1769in,
%eight=1.6385in,
%t={(0in,0in)},
%cale only axis,
%min=0,
%max=1,
%min=0,
%max=1,
%xis line style={draw=none},
%icks=none,
%xis x line*=bottom,
%xis y line*=left
%
%$\end{axis}
\end{tikzpicture}%
    \caption{Behavior is consistent after a reboot of the virtual machine shown in \autoref{fig:tcp_test_stack}, but inconsistent between reboots; this graph shows the likelihood that an IOTLB entry will be consistently evicted by a Prime+Probe after a reboot of the system. Entries marked in red are evicted whether or not there is network activity and do not vary between reboots; entries in blue are those that are evicted when there is network activity but not when there is no activity and vary significantly.}
    \label{fig:variation_between_reboots}
\end{figure}

Prime+Probe data from this experiment are visualized in \autoref{fig:variation_between_reboots}.
There was substantial variation of IOTLB activity after a reboot of the virtual machine operating the Realtek NIC, so results are plotted as means across many reboots.
More evictions were detected in the probes of the Prime+Probe while the network was active, indicating a side-channel leakage in the IOTLB that originated from the Realtek NIC.
There are two other phenomena of note that are observable in the data from this experiment.
First, the excess evictions caused by the network activity (shown in blue in the figure) varied substantially in the number of sets they occupied.
Whenever the virtual machine was rebooted, the number of sets that were evicted during network activity changed, but there were always evictions in one set (set 11).
After examining the network driver source code, we found that it allocates the transaction buffers used by the network card by calling a kernel function \texttt{dma\_map\_single} on startup, and we verified that by unloading and reloading the network driver, we could reproduce the randomizing effect of rebooting the virtual machine.
Second, sets 1-10 and 126-128 were always evicted in the probe, even absent any network activity or with the network drivers unloaded.
This showed that the 128-page eviction sets, while effective in evicting IOTLB entries, were actually bigger than necessary, since they were evicting their own members.

\end{document}